\documentclass[12pt]{article}
\usepackage[dvips]{graphicx}
\usepackage{amsmath}
\usepackage[dvips]{color}

\renewcommand{\bar}[1]{\overline{#1}}
\newcommand{\half}{{$\frac{1}{2}$}} 

\begin{document}

\begin{flushright}
SLAC-PUB-11696 \\
February, 2006
\end{flushright}



\def\bege{\begin{equation}}
\def\ende{\end{equation}}

\def\vdp{{d^{4}p}\over{(2\pi)^{4}}}
\def\vdq{{d^{4}q}\over{(2\pi)^{4}}}
\def\half{{1\over 2}}
\def\slash#1{\;\raise.15ex\hbox{/}\kern-.47em #1}
\def\Dslash{\raise.15ex\hbox{/}\kern-.77em D}
\def\lsim{\mathrel{\mathstrut\smash{\ooalign{\raise2.5pt\hbox{$<$}\cr\lower2.5pt\hbox{$\sim$}}}}}

\def\ddp{d\tilde{p}}
\def\a{\alpha}
\def\b{\beta}
\def\D{\Delta}
\def\G{\Gamma}
\def\e{\epsilon}
\def\g{\gamma}
\def\d{\delta}
\def\p{\phi}
\def\vp{\varphi}
\def\r{\rho}
\def\s{\sigma}
\def\l{\lambda}
\def\L{\Lambda}
\def\th{\theta}
\def\om{\omega}
\def\Om{\Omega}

\def\del{\partial}
\def\ha{\frac{1}{2}}
\def\bar#1{\overline{ #1 }}
\def\psibar{\overline{\psi}}
\def\etabar{\overline{\eta}}
\def\sla#1{#1\!\!\!/}
\def\smgroup{U(1)_Y{\otimes}SU(2)_L{\otimes}SU(3)_C}

\def\ra{\rangle}
\def\la{\langle}
\def\lraw{\leftrightarrow}
\def\raw{\rightarrow}
\def\bdlra{\buildrel \leftrightarrow \over}
\def\bdra{\buildrel \rightarrow \over}

\def\eg{{\it e.g.}}
\def\ie{{\it i.e.}}
\def\Dslash{\raise.15ex\hbox{/}\kern-.77em D}
\def\Aslash{\raise.15ex\hbox{/}\kern-.77em A}
\def\Hslash{\raise.15ex\hbox{/}\kern-.77em H}
\def\np#1#2#3{Nucl. Phys. {\bf #1} (#2) #3}
\def\pl#1#2#3{Phys. Lett. {\bf #1} (#2) #3}
\def\prl#1#2#3{Phys. Rev. Lett. {\bf #1} (#2) #3}
\def\pr#1#2#3{Phys. Rev. {\bf #1} (#2) #3}
\def\prd#1#2#3{Phys. Rev. D {\bf #1} (#2) #3}
\def\etal{{\it et al}}

\def\ddp{d\tilde{p}}
\def\ddl{d\tilde{l}}
\def\ddk{d\tilde{k}}
\def\g{\gamma}
\def\mm{m_{\pi}}
\def\ra{\rangle}
\def\la{\langle}
\def\gt{\tilde{\gamma}}
\def\mQbar{m_{\bar{Q}}}
\def\Lraw{\Longrightarrow}
\def\mgx{m_{\tilde{g}}}
\def\ggx{\tilde{g}}
\def\mqx{m_{\tilde{q}}}
\def\qx{\tilde{q}}
\def\mlx{m_{\tilde{l}}}
\def\lx{\tilde{l}}
\def\wx{\tilde{w}}
\def\tx{\tilde{t}}
\def\bx{\tilde{b}}
\def\hpm{{H}^{\pm}}
\def\hpmx{\tilde{H}^{\pm}}

\def\lb{\bar{\lambda}}
\def\ppp{(\!+\!+\!+\!)}
\def\ppm{(\!+\!+\!-\!)}
\def\pmp{(\!+\!-\!+\!)}
\def\mpp{(\!-\!+\!+\!)}
\def\pmm{(\!+\!-\!-\!)}
\def\mpm{(\!-\!+\!-\!)}
\def\mmp{(\!-\!-\!+\!)}
\def\mmm{(\!-\!-\!-\!)}
\def\CQ{{\cal Q}}
\def\CK{{\cal K}}
\def\CP{{\cal P}}
\def\CL{{\cal L}}
\def\pot{(p_1\!\cdot\! p_2)}
\def\ptt{(p_2\!\cdot\! p_3)}
\def\pto{(p_3\!\cdot\! p_1)}

\def\Li2{{\rm Li}_2}
\def\Cl2{{\rm Cl}_2}
\def\Clh2{{\rm Clh}_2}
\def\AClh2{{\rm AClh}_2}

\def\gG{{\cal G}}
\def\O{{\cal O}}
\def\M{{\cal M}}
\def\tilde{\widetilde}
\def\bar{\overline}
\def\Z{{\bf Z}}
\def\T{{\bf T}}
\def\S{{\bf S}}
\def\R{{\bf R}}
\def\np#1#2#3{Nucl. Phys. {\bf B#1} (#2) #3}
\def\pl#1#2#3{Phys. Lett. {\bf #1B} (#2) #3}
\def\prl#1#2#3{Phys. Rev. Lett.{\bf #1} (#2) #3}
\def\physrev#1#2#3{Phys. Rev. {\bf D#1} (#2) #3}
\def\ap#1#2#3{Ann. Phys. {\bf #1} (#2) #3}
\def\prep#1#2#3{Phys. Rep. {\bf #1} (#2) #3}
\def\rmp#1#2#3{Rev. Mod. Phys. {\bf #1} (#2) #3}
\def\cmp#1#2#3{Comm. Math. Phys. {\bf #1} (#2) #3}
\def\mpl#1#2#3{Mod. Phys. Lett. {\bf #1} (#2) #3}

\def\Lam#1{\Lambda_{#1}}
\def\pf{{\rm Pf ~}}
\font\zfont = cmss10 
\font\litfont = cmr6
\font\fvfont=cmr5
\def\bigone{\hbox{1\kern -.23em {\rm l}}}
\def\ZZ{\hbox{\zfont Z\kern-.4emZ}}
\def\half{{\litfont {1 \over 2}}}
\def\mx#1{m_{\hbox{\fvfont #1}}}
\def\gx#1{g_{\hbox{\fvfont #1}}}
\def\lamlam#1{\langle S_{#1}\rangle}
\def\Re{{\rm Re ~}}
\def\Im{{\rm Im ~}}
\def\lfm#1{\medskip\noindent\item{#1}}




\bigskip\bigskip

\begin{center}
{\Large \bf The Form Factors of the Gauge-Invariant Three-Gluon
Vertex\footnote{Work supported by the Department of Energy under
contract number DC-AC02-76SF00515}}
\end{center}

\vspace{13pt}

\centerline{ \bf Michael Binger and Stanley J. Brodsky}

\vspace{8pt} {\centerline{Stanford Linear Accelerator Center,}}

{\centerline{Stanford University, Stanford, California 94309, USA}}

\centerline{e-mail: binger@slac.stanford.edu and sjbth@slac.stanford.edu}

\bigskip\bigskip

\begin{abstract}

The gauge-invariant three-gluon vertex obtained from the pinch
technique is characterized by thirteen nonzero form factors, which
are given in complete generality for unbroken gauge theory at one
loop. The results are given in $d$ dimensions using both dimensional
regularization and dimensional reduction, including the effects of
massless gluons and arbitrary representations of massive gauge
bosons, fermions, and scalars. We find interesting relations between
the functional forms of the contributions from gauge bosons,
fermions, and scalars. These relations hold only for the
gauge-invariant pinch technique vertex and are d-dimensional
incarnations of supersymmetric nonrenormalization theorems which
include finite terms. The form factors are shown to simplify for
$N=1,2$, and $4$ supersymmetry in various dimensions. In
four-dimensional non-supersymmetric theories, eight of the form
factors have the same functional form for massless gluons, quarks,
and scalars, when written in a physically motivated tensor basis.
For QCD, these include the tree-level tensor structure which has
prefactor $\beta_0=(11N_c-2N_f)/3$, another tensor with prefactor
$4N_c-N_f$, and six tensors with $N_c-N_f$. In perturbative
calculations our results lead naturally to an effective coupling for
the three-gluon vertex, $\tilde{\a}(k_1^2,k_2^2,k_3^2)$, which
depends on three momenta and gives rise to an effective scale
$Q_{eff}^2(k_1^2,k_2^2,k_3^2)$ which governs the behavior of the
vertex. The effects of nonzero internal masses $M$ are important and
have a complicated threshold and pseudo-threshold structure. A
three-scale effective number of flavors
$N_F(k_1^2/M^2,k_2^2/M^2,k_3^2/M^2)$ is defined. The results of this
paper are an important part of a gauge-invariant dressed skeleton
expansion and a related multi-scale analytic renormalization scheme.
In this approach the scale ambiguity problem is resolved since
physical kinematic invariants determine the arguments of the
couplings.
\end{abstract}

\newpage



\section{Introduction: Gauge-Invariant Green's Functions}

The main purpose of this paper is to analyze the structure of the
gauge-invariant three-gluon vertex \cite{Cornwall:1989gv},
calculate the fourteen form factors at one loop, and outline some
of the phenomenological applications. Before proceeding, it is
worthwhile to review the motivation and current status of
gauge-invariant Green's functions.

In the conventional formulation of gauge field theories, the
manifest gauge-invariance of the original action is lost upon
quantization, simply because one has to fix a gauge in order to
perform calculations. Generically, Greens's functions are
gauge-dependent and thus not physical by themselves. Only the
particular combinations of Green's functions which form physical
observables must be gauge-invariant. In many theoretical studies,
however, one would like to consider individual Green's functions and
extract physical meaning from them \cite{Cornwall:1981zr}. For
example, studies of the infrared behavior of gauge theory using
Dyson-Schwinger equations \cite{Alkofer:2000wg} often rely on
gauge-dependent truncation schemes which one hopes are not too
brutal. The existence of gauge-invariant two-point functions is
crucial for defining meaningful resummed propagators
\cite{Papavassiliou:1996zn}, particularly near threshold, for the
construction of effective charges \cite{Watson:1996fg}, for a
postulated dressed-skeleton expansion of QCD \cite{Brodsky:2000cr},
and for justifying renormalon analyses \cite{Beneke:1998ui}.

 Thus there is strong motivation for gauge-invariant Green's
functions with physical content. We will now briefly discuss the
relationship between three different approaches to gauge-invariant
Green's functions : (1) the Pinch Technique (PT), (2) the Background
Field Method (BFM), and (3) the $\star$ effective Lagrangian scheme
of Kennedy and Lynn. All three approaches will lead to the same
Green's functions.

 The pinch technique (PT) was first constructed by Cornwall
\cite{Cornwall:1981zr} in order to study gauge-invariant
Dyson-Schwinger equations and dynamical gluon mass generation, but
the approach is much more generally applicable. In the PT approach,
unique gauge-invariant Green's functions are constructed by
explicitly rearranging Feynman diagrams using elementary Ward
identities (WI) as the guiding principle. Longitudinal momenta from
triple-gauge-boson vertices and gauge propagators inside of loops
hit other vertices and thus generate inverse propagators (via WI's),
which, in turn, cancel (or pinch) some internal propagators. In this
way, certain parts of Green's functions are reduced to parts of
lower $n$-point functions, and should properly be included in the
latter.

As an example of the PT, consider the gluon (or massive gauge boson)
self-energy. The conventional self-energy is gauge-dependent and
physically meaningless by itself. However, when embedded in any
physical process, there will be associated parts of vertex and box
graphs which undergo the reduction described above and thus have the
same tensor and kinematic structure as the gluon propagator. These
pinched parts are then added to the conventional gauge-dependent
self-energy, yielding a gauge-invariant self-energy and gluon
propagator that has the correct asymptotic UV behavior dictated by
the renormalization group equation. The resulting two-point function
has numerous positive attributes
\cite{Papavassiliou:1996zn}\cite{Degrassi:1992ue}\cite{Watson:1996fg}\cite{Philippides:1995gf}\cite{Papavassiliou:1994fp},
including uniqueness, resummability, analyticity, unitarity, and a
natural relation to optical theorem, from which it can also be
derived \cite{Papavassiliou:1996zn}\cite{Papavassiliou:1996fn}.

Resumming these two-point functions leads to physical effective
charges, $\acute{a} la$ Grunberg\cite{Grunberg:1982fw}, which can be
extended to the supersymmetric case and leads to an analytic
improvement of gauge coupling unification with smooth threshold
behavior \cite{Binger:2003by}.

This method has been applied to a variety of Green's functions
\cite{Cornwall:1989gv}\cite{Papavassiliou:1995fw}\cite{Papavassiliou:1989zd}
\cite{Papavassiliou:1994pr}\cite{Papavassiliou:1992ia}, with
applications to electroweak phenomenology
\cite{Bernabeu:2002nw}\cite{Papavassiliou:1993ex}. In particular,
the gauge-invariant three-gluon vertex was first constructed in
\cite{Cornwall:1989gv} to one-loop order, where the authors showed
that the vertex satisfies a relatively simple abelian-like Ward
identity. However, the integrals were not evaluated, so that little
could be said about the individual form factors except that the UV
divergent term in the tree level tensor structure is correct. The
main motivation of this paper is to extend this work by evaluating
the integrals for the fourteen form factors, and expressing the
results in a convenient tensor basis for phenomenological
applications. In doing so, an interesting structure emerges, in
which the contributions of gluons(G), quarks(Q), and scalars(S) are
intimately related. These relations are closely linked to
supersymmetry and conformal symmetry, and in particular the $N=4$
non-renormalization theorems. For all form factors $F$ in dimensions
$d$, we find that
 \bege
 F_G+4F_Q+(10-d)F_S=0,
 \ende
which encodes the vanishing contribution of the $N=4$ supermultiplet
in four dimensions. Similar relations have been found in the context
of supersymmetric scattering amplitudes
\cite{Bern:1994zx}\cite{Bern:1994cg}. In Appendix E, the effects of
internal masses are discussed, and the above sum rule becomes
modified
 \bege
 F_{MG}+4F_{MQ}+(9-d)F_{MS}=0,
 \ende
for internal massive gauge bosons (MG), fermions (MQ), and scalar
(MS). The external gluons remain massless and unbroken, so the
internal gauge bosons might be the heavy $X_{\mu},Y_{\mu}$ bosons of
$SU(5)$, for example. In \cite{Bern:1993tz}, supersymmetric
relations were found for electroweak gauge boson four-point
scattering amplitudes.

The PT method has been explicitly extended beyond one-loop
\cite{Papavassiliou:1999az}\cite{Papavassiliou:1999bb}\cite{Binosi:2001hy},
has recently been proven to exist to all orders in perturbation
theory
\cite{Binosi:2002ft}\cite{Binosi:2002vk}\cite{Binosi:2004qe}\cite{Binosi:2003rr},
and interestingly, each Green's function is equal to the
corresponding Green's function of the Background Field Method
(BFM) in quantum Feynman gauge $\xi_Q=1$, a result suggested in
\cite{Denner:1994nn}\cite{Papavassiliou:1994yi}. Heuristically,
this is due to the fact that there are no longitudinal (pinching)
momenta in the gauge propagator or the elementary vertices in this
special gauge.

 The Background Field Method (BFM) \cite{bfm} constructs manifestly gauge invariant Green's
 functions in the following way. First, the field
 variable($A$) in the path integral is separated into a background($B$) and
 quantum($Q$) field, $A=B+Q$. Only the quantum field $Q$ propagates in loops, since it is
 a variable of functional integration. In contrast, the background field $B$
 appears only in external legs. By judiciously
 choosing the gauge-fixing function, one arrives at an effective
 action which remains manifestly invariant under background field
 gauge-transformations
 $\d B^a_{\mu} = -f^{abc}\omega^bB^c_{\mu}+{1\over g}\del_{\mu}\omega^a$.
Furthermore, derivatives of the BFM effective action with respect to
the background field $B$ yield the same 1PI Green's functions as the
conventional effective action with a nonstandard gauge-fixing. Thus,
it can be shown \cite{bfm} that the correct S-matrix is obtained by
sewing together trees composed of 1PI Green's functions of $B$
fields. In doing so, one can fix the gauge of $B$, which propagates
only at tree level, independently of the gauge fixing of $Q$. For
example, convenient non-covariant gauges might be used for the trees
while BFM Feynman gauge $\xi_Q=1$ (BFMFG) can be used for the loops.

The correspondence between the PT and BFM is not surprising, since
the BFM is a formulation of gauge theory where Green's functions of
the gauge field are manifestly (background) gauge-invariant.
Although this is true for all values of the quantum gauge-fixing
parameter $\xi_Q$, it is only for the special value $\xi_Q=1$ that
the BFM Green's functions also have the correct kinematic structure
of the irreducible PT Green's functions. Alternatively, it has been
shown \cite{Papavassiliou:1994yi} that applying the PT algorithm to
the BFM for $\xi_Q\neq 1$ leads back to the canonical ($\xi_Q=1$) PT
Green's functions.

Finally, in the $\star$ scheme of Kennedy and Lynn
\cite{Kennedy:1988sn}, a gauge-invariant effective Lagrangian was
constructed for electroweak four-fermion processes by explicitly
rearranging the one loop corrections. As in the pinch technique,
vertex parts must be added to would-be two point functions to
yield genuine two-point functions. One particular motivation is
that fact that the photon acquires a spurious mass from its mixing
with $Z^0$, $\Pi_{\g Z}(q^2=0)\neq 0$, unless the correct vertex
parts are added. The resulting effective charges, $\a_\star(q^2)$
and $s^2_\star(q^2)$ are in fact precisely equal to the
corresponding pinch-technique effective charges at one loop,
including all finite terms and threshold dependence
\cite{Degrassi:1992ue}.

Thus, all three methods for constructing physical gauge-invariant
Green's functions lead to the same results, which in the this paper
will be referred to as either PT or PT/BFMFG Green's functions.

 The organization of this paper is as follows. In section
2, we will discuss the general structure of the gauge-invariant
three-gluon vertex, which is constrained by the Ward identity and
Bose symmetry. Two convenient tensor bases and their relation are
discussed.  In section 3, the main results of this paper are given.
First, the nontrivial supersymmetric relations between the gluon,
quark, and scalar contributions to each form factor are discussed.
The explicit results for the form factors are given in two different
bases for massless internal particles, with the full mass effects
relegated to Appendix E. In section 4, we briefly discuss the
phenomenological application to physical scattering processes, where
we derive an effective coupling for the three-gluon vertex,
$\tilde{\a}(k_1^2,k_2^2,k_3^2)$, and an effective scale,
$Q^2_{eff}(k_1^2,k_2^2,k_3^2)$, both of which depend on three
distinct gluon virtualities. In section 5, the phenomenological
effects of internal masses are discussed. A complicated threshold
and pseudo-threshold structure emerges. Furthermore, a three-scale
effective number of flavors $N_F(k_1^2/M^2,k_2^2/M^2,k_3^2/M^2)$ is
defined. Conclusions are given in section 6. In appendix A, a brief
outline of the calculational method is given, and some basic one-
and two-point integrals are given. Appendix B is devoted to a
thorough discussion of the massive triangle integral, and analytic
continuations are given for each kinematic region. Appendix C
collects some useful results for special functions. Appendix D
explains the corrections to the form factors when a supersymmetric
regularization is used. Finally Appendix E gives explicitly the
corrections to the form factors arising from internal massive gauge
bosons, fermions, and scalars.

\section{General Structure of the Three-Gluon Vertex}

\subsection{Symmetries}

One of the most important aspects of the gauge-invariant three-gluon
vertex discussed in this paper is the relatively simple Ward
identity it satisfies, which has the same form as the Ward ID
satisfied by the tree level vertex. This was proven at one-loop in
the original paper by Cornwall and Papavassiliou
\cite{Cornwall:1989gv} using the explicit one-loop result, which is
the gluon part of Eqs.(\ref{fullvertex},\ref{GQSdef}) below. It is
straightforward to show that the fermion and scalar parts also
satisfy the same Ward identity (just as in QED). Furthermore, the
equivalence of the BFMFG and PT to all orders \cite{Binosi:2002ft}
allows one to write the Ward identity satisfied by the three-gluon
vertex to all orders as
 \bege\label{wardid}
p_3^{\mu_3}\Gamma_{\mu_1\mu_2\mu_3}^{abc}(p_1,p_2,p_3) =
 f^{adc}\Big( t_{\mu_1\mu_2}(p_2)  \d^{db} +
 \Pi_{\mu_1\mu_2}^{db}(p_2)\Big)
 -f^{adc}\Big( t_{\mu_1\mu_2}(p_1)  \d^{da} +
 \Pi_{\mu_1\mu_2}^{da}(p_1)\Big),
 \ende
plus two other equations which are cyclic permutations. The
transverse tensor $t_{\mu\nu}(p)=p^2g_{\mu\nu}-p_\mu p_\nu$ comes
from the tree level term. Here all momenta are defined to be
incoming and all labels are defined in counter-clockwise fashion, as
shown in Fig. 1. This Ward identity represents a great
simplification compared to the usual Slavnov-Taylor identities
satisfied by the conventional gauge-dependent three-gluon vertex,
which involves the gluon propagator, the ghost propagator, and the
ghost-ghost-gluon vertex function. The self-energy function in the
above equation is not the usual gauge-dependent self-energy, but
rather the gauge-invariant pinch technique self energy, which is the
only self energy discussed in this paper.
\begin{figure}[htb]
 \centering \includegraphics[height=2.5in]{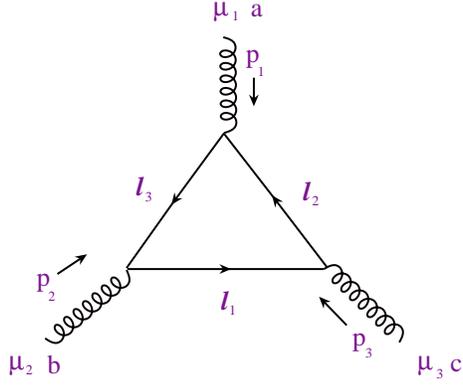} \caption[*]{The notation
 and loop momentum routing used throughout this paper. The internal particle could be
 a gauge boson, ghost, quark, or scalar.} \label{fig:1}
\end{figure}
An immediate consequence is that the longitudinal ($L$) part of the
vertex, defined as the part which contributes to the above Ward ID,
must have only the antisymmetric color factor $f^{abc}$ so long as
gluons conserve the color charge, $\Pi_{\mu\nu}^{ab}(q) =
\d^{ab}\Pi_{\mu\nu}(q)$. As far as we know, the transverse ($T$)
part of the vertex (defined by
$p_3^{\mu_3}\Gamma_{\mu_1\mu_2\mu_3}^{abc (T)}(p_1,p_2,p_3) =0$) is
not required to be proportional to $f^{abc}$, but may in principle
contain $d^{abc}$ terms. Nevertheless, no such terms appear at one
or two loop order, and so in the subsequent discussion we take
 \bege
\Gamma_{\mu_1\mu_2\mu_3}^{abc}(p_1,p_2,p_3) =
f^{abc}\Gamma_{\mu_1\mu_2\mu_3}(p_1,p_2,p_3),
 \ende
in terms of which the Ward identity becomes
 \bege\label{wardid2}
 p_3^{\mu_3}\Gamma_{\mu_1\mu_2\mu_3}(p_1,p_2,p_3) =
  t_{\mu_1\mu_2}(p_2) (1+\Pi(p_2^2))
 -t_{\mu_1\mu_2}(p_1) (1+\Pi(p_1^2)).
 \ende

Bose symmetry, the fact that 3 identical particles are entering the
vertex, and the properties of $f^{abc}$ imply definite properties of
$\Gamma_{\mu_1\mu_2\mu_3}(p_1,p_2,p_3)$ under the interchange of
labels.  In particular, defining the five elements of the
permutation group ${\cal S}_3$ to act by
 \begin{eqnarray}\label{gops}
 g_{123} &=&\left( \begin{array}{ccc}
 (\mu_1,p_1)\raw(\mu_2,p_2) \\
 (\mu_2,p_2)\raw(\mu_3,p_3) \\
 (\mu_3,p_3)\raw(\mu_1,p_1)
 \end{array} \right)
 \;\;\;\;\;\;\;\;
 g_{12} = \left( \begin{array}{ccc}
 (\mu_1,p_1)\raw(\mu_2,p_2) \\
 (\mu_2,p_2)\raw(\mu_1,p_1)
 \end{array} \right)
 \\
 g_{23} &=& \left( \begin{array}{ccc}
 (\mu_2,p_2)\raw(\mu_3,p_3) \\
 (\mu_3,p_3)\raw(\mu_2,p_2)
 \end{array} \right)
 \;\;\;\;\;\;\;\;
 g_{31} =\left( \begin{array}{ccc}
 (\mu_3,p_3)\raw(\mu_1,p_1) \\
 (\mu_1,p_1)\raw(\mu_3,p_3)
 \end{array} \right)
\nonumber
\end{eqnarray}
and $g_{321}=g_{123}^{-1}$ one finds that
$(g_{123},g_{321},g_{12},g_{23},g_{31})$ yields $(+,+,-,-,-)$ when
acting on $\Gamma_{\mu_1\mu_2\mu_3}(p_1,p_2,p_3)$.

The nonabelian nature of the permutation group ${\cal S}_3$ prevents
one from finding a basis in which all of the tensors are eigenstates
of all of these operators. Thus, aesthetic and physical principles
must guide us in choosing convenient bases, two of which are
discussed momentarily.

\subsection{Two Convenient Bases}

 Let us consider the most general tensor structure. The three index Lorentz covariant
 tensor must be constructed out of the metric $g_{\mu\nu}$ and the
 momenta ($p_i^{\mu}$). Since the momenta are not independent,
 $p_1+p_2+p_3=0$, simple combinatorics implies that there are in
 general 14 independent tensor components, 6 of which have one power
 of momenta and also the metric, and 8 of which have 3 powers of
 momenta. Many different basis choices can be made, although we will use essentially two.

 In the subsequent discussion, some efficient notation will prove
 useful. This is summarized in Table 1.

 $$
 {\rm Table \;\; I. \; Definition \;\; of \;\; tensor\;\; abbreviations}
 $$
 $$
 \label{bh}
 \def\vspace#1{ \omit & \omit & height #1 &  \omit && \omit &\cr }
 \vbox{\offinterlineskip
 \hrule
 \halign{
 \strut
 \vrule#& $\;$ \hfil # \hfil &
 \vrule#& $\;$ \hfil # \hfil &
 \vrule# \cr
 \noalign{\hrule}
 &  $ 001 \equiv g_{\mu_1\mu_2}p_{1\mu_3} $
 && $ 002 \equiv g_{\mu_1\mu_2}p_{2\mu_3} $ &\cr
\noalign{\hrule}
 &  $ 200 \equiv g_{\mu_2\mu_3}p_{2\mu_1} $
 && $ 300 \equiv g_{\mu_2\mu_3}p_{3\mu_1} $ &\cr
\noalign{\hrule}
 &  $ 030 \equiv g_{\mu_3\mu_1}p_{3\mu_2} $
 && $ 010 \equiv g_{\mu_3\mu_1}p_{1\mu_2} $ &\cr
\noalign{\hrule}
 &  $ 211 \equiv p_{2\mu_1}p_{1\mu_2}p_{1\mu_3} $
 && $ 212 \equiv p_{2\mu_1}p_{1\mu_2}p_{2\mu_3} $ &\cr
\noalign{\hrule}
 &  $ 232 \equiv p_{2\mu_1}p_{3\mu_2}p_{2\mu_3} $
 && $ 332 \equiv p_{3\mu_1}p_{3\mu_2}p_{2\mu_3} $ &\cr
\noalign{\hrule}
 &  $ 331 \equiv p_{3\mu_1}p_{3\mu_2}p_{1\mu_3} $
 && $ 311 \equiv p_{3\mu_1}p_{1\mu_2}p_{1\mu_3} $ &\cr
\noalign{\hrule}
 &  $ 312 \equiv p_{3\mu_1}p_{1\mu_2}p_{2\mu_3} $
 && $ 231 \equiv p_{2\mu_1}p_{3\mu_2}p_{1\mu_3} $ &\cr
 }
 \hrule}
 $$

Thus, each tensor is rewritten as a 3 slot object, where slots
correspond to $\mu_1,\mu_2,\mu_3$ in that order, and the content of
each slot is either `1',`2', or `3' to represent momentum
$p_1,p_2,p_3$, or a `0', which must occur in pairs and represents
that those two indices are connected by the metric tensor.

The most naive thing to do would be to just eliminate one momenta,
say $p_3=-p_1-p_2$ and use the following 14 basis tensors : $100$,
$200$, $010$, $020$, $001$, $002$, $111$, $112$, $121$, $211$,
$122$, $212$, $221$, $222$. This is not very useful since the
explicit Bose symmetry between the three gluons has been broken, and
thus delicate relations between the form factors will have to
enforce it.

{\center{\Large \bf The $\pm$ Basis}}

A more natural choice is obtained by starting from a manifestly
symmetric, but redundant basis, which has 36 possible basis tensors,
9 with one power of momenta, and 27 with three powers of momenta. As
a step towards our final basis, we find it convenient to eliminate
all such tensors with momenta $p_1^{\mu_1}, p_2^{\mu_2}$, or
$p_3^{\mu_3}$, i.e. anything with 1 in the first slot, 2 in second
slot or 3 in the third slot. This yields the 14 basis tensors $001$,
$002$, $200$, $300$, $030$, $010$, $211$, $212$, $232$, $332$,
$331$, $311$, $231$, $312$, which are shown in Table 1. Note that
under the action of $g_{123}$ we have $200\raw 030, 300\raw 010,
211\raw 232, 311\raw 212$, etc. Also, notice that $200$ and $300$
are interchanged by the action of $g_{23}$, while $211$ and $212$
are interchanged by the action of $g_{12}$, etc. Thus, it is
convenient to take appropriate linear combinations such that one of
these interchange operators is diagonal for each tensor. Such basis
tensors are
 \begin{eqnarray}\label{pmbasis}
  \hat{a}_{12}  &=& (00-) = 001 - 002, \;\;\; \hat{a}_{23} = (-00) = 200 - 300, \;\;\; \hat{a}_{31} = (0\!-\!0) = 030 - 010\nonumber\\
  \hat{b}_{12}  &=& (00+) = 001 + 002, \;\;\; \hat{b}_{23} = (+00) = 200 + 300, \;\;\; \hat{b}_{31} = (0\!+\!0) = 030 + 010\nonumber\\
  \hat{c}_{12}  &=& (\!+\!+\!-\!), \;\;\; \hat{c}_{23} = (\!-\!+\!+\!), \;\;\; \hat{c}_{31} = (\!+\!-\!+\!) \nonumber\\
  \hat{d}_{12}  &=& (\!-\!-\!+\!), \;\;\; \hat{d}_{23} = (\!+\!-\!-\!), \;\;\; \hat{d}_{31} = (\!-\!+\!-\!) \\
  \hat{h} &=& (\!-\!-\!-\!), \;\;\; \hat{s} = (\!+\!+\!+\!)\nonumber,
 \end{eqnarray}
 where the notation means $(\!\pm\!\pm\!\pm\!)\;\equiv\;(2\pm3,3\pm1,1\pm2)$, so that $(\!+\!+\!-\!) =
 231-232+211-212+331-332+311-312$, etc.
 The subscripts are chosen because, for example, $\hat{a}_{12}$ is an
 eigenstate of $g_{12}$, etc.

Suppressing indices and momentum dependence, the three-gluon vertex
is then written as
 \begin{eqnarray}\label{3gvertex}
 \Gamma = (A_{12}\hat{a}_{12} + B_{12}\hat{b}_{12}+C_{12}\hat{c}_{12} + D_{12}\hat{d}_{12} + {\rm perms})
  + S\hat{s} + H\hat{h},
 \end{eqnarray}
where the lower case letters represent the basis tensors, while the
upper case letters are the form factors, which depend on $p_1^2,
p_2^2,$ and $p_3^2$. In addition to indicating which basis tensors
they are associated with, the subscripts on form factors also
indicate the ordering of momenta in the arguments. For example,
$A_{12} = A(p_1^2,p_2^2|p_3^2)$, $A_{23} = A(p_2^2,p_3^2|p_1^2)$,
$A_{31} = A(p_3^2,p_1^2|p_2^2)$, and the first two arguments are
either symmetric or antisymmetric. The behavior of these form
factors under ${\cal S}_3$ can be inferred from the behavior of the
basis tensors under ${\cal S}_3$ (which will be discussed
momentarily), along with the overall requirement for the vertex
given below Eq. \ref{gops}. One finds that $A(x,y|z)=+A(y,x|z)$, and
thus $A_{12}=A_{21}$, etc. Similarly, $B_{12}=-B_{21}$,
$C_{12}=C_{21}$, and $D_{12}=-D_{21}$. $H$ is totally invariant
under the interchange or permutation of any momenta, while $S$ goes
to $-S$ under any interchange of momenta, but is invariant under a
cyclic permutation $g_{123}$.

It is straightforward to see that under the action of the
permutation operator ($g_{123}$) these fourteen basis tensors are
organized into four triplets,
$\{\hat{a}_{12},\hat{a}_{23},\hat{a}_{31}\}$,
$\{\hat{b}_{12},\hat{b}_{23},\hat{b}_{31}\}$,
$\{\hat{c}_{12},\hat{c}_{23},\hat{c}_{31}\}$,
$\{\hat{d}_{12},\hat{d}_{23},\hat{d}_{31}\}$, as well as $\hat{h}$
and $\hat{s}$. The latter two are eigenstates of all five operators.

 Consider the properties of
$\{\hat{a}_{12},\hat{a}_{23},\hat{a}_{31}\}$ under the permutation
group. It is easy to see that under the action of any element $g_i$,
we have \bege
 \left( \begin{array}{cccc}
 \hat{a}_{12} \\
 \hat{a}_{23} \\
 \hat{a}_{31}
 \end{array} \right)
 \raw g_i \left( \begin{array}{cccc}
 \hat{a}_{12} \\
 \hat{a}_{23} \\
 \hat{a}_{31}
 \end{array} \right),
 \ende
with the matrices given by
 \begin{eqnarray}\label{gmats}
 &g_{123}& =\left( \begin{array}{cccc}
 0 & 1 & 0 \\
 0 & 0 & 1 \\
 1 & 0 & 0
 \end{array} \right)\;\;\;\;\;
 g_{321} =\left( \begin{array}{cccc}
 0 & 0 & 1 \\
 1 & 0 & 0 \\
 0 & 1 & 0
 \end{array} \right) \\
 &g_{12}& = -\left( \begin{array}{cccc}
 1 & 0 & 0 \\
 0 & 0 & 1 \\
 0 & 1 & 0
 \end{array} \right)\;\;\;\;
 g_{23} = -\left( \begin{array}{ccc}
 0 & 0 & 1 \\
 0 & 1 & 0 \\
 1 & 0 & 0
 \end{array} \right)\;\;\;\;
 g_{31} = -\left( \begin{array}{ccc}
 0 & 1 & 0 \\
 1 & 0 & 0 \\
 0 & 0 & 1
 \end{array} \right)\nonumber
\end{eqnarray}

The transformation rules are identical for $\{c_{ij}\}$, and similar
for $\{b_{ij}\}$ and $\{d_{ij}\}$ with the only change being that
there is no minus sign in the three interchange operators $g_{12},
g_{23}$ and $g_{31}$.

The basis constructed above (Eq.(\ref{pmbasis})) will be called the
$\pm$ basis. As discussed later, this basis is the most convenient
for phenomenology and furthermore the form factors exhibit
particularly simple relations between the gluon, quark, and scalar
contributions (Eqs.(\ref{QGsumpm},\ref{QGsumsym})).

However, the $\pm$ basis as it stands does not contain the tree
level tensor structure. Thus, one is naturally led to diagonalizing
the permutation operator $g_{123}$.\footnote{One can readily check
that the only two operators in ${\cal S}_3$ which commute are
$g_{123}$ and its inverse $g_{321}$. Thus one can diagonalize these
two, OR one of the interchange operators $g_{12},g_{23},g_{31}$.}
Clearly, this is the most symmetric choice and, more importantly,
one of the resulting eigenvectors is the tree level tensor
structure.

In the triplet representation of ${\cal S}_3$, $g_{123}$ is
diagonalized by the similarity transformation
 \bege
S = {1\over \sqrt{3}}\left( \begin{array}{ccc}
 1 & 1 & 1 \\
 1 & \l & \lb \\
 1 & \lb & \l
 \end{array} \right)
 \;\;\;\;\;\;\;\;
 S^{-1}g_{123}S \;\equiv\; \tilde{g}_{123} = \left( \begin{array}{ccc}
 1 & 0 & 0 \\
 0 & \l & 0 \\
 0 & 0 & \lb
 \end{array} \right),
 \ende
 where $\l=\exp{({2i\pi\over 3})}= -\half+i{\sqrt{3}\over 2}$, $\lb=\l^*$ are cube roots of
 unity.
This results in new basis tensors and form factors
\begin{eqnarray}\label{sympmdef}
\left( \begin{array}{ccc}
 \hat{a}_0 \\
 \hat{a}_+ \\
 \hat{a}_-
 \end{array} \right)
 &\equiv& \sqrt{3} S^{-1}
 \left( \begin{array}{ccc}
 \hat{a}_{12} \\
 \hat{a}_{23} \\
 \hat{a}_{31}
 \end{array} \right)
 = \left( \begin{array}{ccc}
 1 & 1 & 1 \\
 1 & \lb & \l \\
 1 & \l & \lb
 \end{array} \right)
 \left( \begin{array}{ccc}
 \hat{a}_{12} \\
 \hat{a}_{23} \\
 \hat{a}_{31}
 \end{array} \right)
 \nonumber\\
\left( \begin{array}{ccc}
 A_0 \\
 A_+ \\
 A_-
 \end{array} \right)
 &\equiv& {1\over\sqrt{3}} S
 \left( \begin{array}{ccc}
 A_{12} \\
 A_{23} \\
 A_{31}
 \end{array} \right)
 = {1\over 3}\left( \begin{array}{ccc}
 1 & 1 & 1 \\
 1 & \l & \lb \\
 1 & \lb & \l
 \end{array} \right)
 \left( \begin{array}{ccc}
 A_{12} \\
 A_{23} \\
 A_{31}
 \end{array} \right).
\end{eqnarray}
This procedure is repeated identically for the $(b,B)$, $(c,C)$, and
$(d,D)$ basis tensors and form factors.

Notice that $\hat{a}_0 = g_{\mu_1\mu_2}(p_1-p_2)_{\mu_3} +
g_{\mu_2\mu_3}(p_2-p_3)_{\mu_1} + g_{\mu_3\mu_1}(p_3-p_1)_{\mu_2}$
is the tree level tensor, which is why the extra factors of
$\sqrt{3}$ were included above.

The transformation properties for the basis tensors $\hat{a}_i$ and
$\hat{c}_i$ are deduced from $\tilde{g}_{123}$ given above,
$\tilde{g}_{321} = {\rm diag}(1,\lb,\l)$, and
 \bege
 \tilde{g}_{12} =-\left( \begin{array}{cccc}
 1 & 0 & 0 \\
 0 & 0 & 1 \\
 0 & 1 & 0
 \end{array} \right)
 \;\;\;\;
 \tilde{g}_{23} = -\left( \begin{array}{ccc}
 1 & 0 & 0 \\
 0 & 0 & \l \\
 0 & \lb & 0
 \end{array} \right)
 \;\;\;\;
 \tilde{g}_{31} = -\left( \begin{array}{ccc}
 1 & 0 & 0 \\
 0 & 0 & \lb \\
 0 & \l & 0
 \end{array} \right),
 \ende
while the transformation properties of the form factors are deduced
by demanding that behavior given below Eq.(\ref{gops}) is respected.
For example, since $\hat{a}_- \raw -\lb \hat{a}_+$ under
$\tilde{g}_{23}$, we find $ A_- \raw \l A_+$ so that $A_-\hat{a}_-
\raw - A_-\hat{a}_-$. For $(b,B)$ and $(d,D)$, the only change in
the above is that there is not a minus sign in
$\tilde{g}_{12},\tilde{g}_{23},$ and $\tilde{g}_{31}$.

We have not touched $\hat{h}$ and $\hat{s}$, as these are already
eigenstates of all five operators
$(g_{123},g_{321},g_{12},g_{23},g_{31})$, with eigenvalues $(++---)$
and $(+++++)$, respectively.

We will call the above constructed basis the symmetric $\pm$ basis.
Note that any basis can be symmetrized in the same manner by
diagonalizing the permutation operator $g_{123}$. The basis we
started with in Eq.(\ref{pmbasis}) is motivated by (a) its simple
and symmetric construction from only metrics and
 $(p_{i+1} {\pm} p_{i-1})_{\mu_i}$, (b) it is the most convenient
 basis for perturbative calculations, as will be discussed in section 4,
 and (c) The individual form factors have a relatively simple
 form, as will be discussed in section 3.

{\center{\Large \bf The LT Basis}}

 For some theoretical studies, another convenient basis is
determined by the distinction between transverse ($T$) and
longitudinal ($L$) tensors \cite{Ball}\cite{Davydychev:1996pb}. The
$L$ tensors contribute to the Ward ID (or the more complicated
Slavnov-Taylor ID for the gauge dependent vertex) while the $T$
tensors satisfy homogeneous equations
$p_3^{\mu_3}\Gamma^{(T)}_{\mu_1\mu_2\mu_3}=0$. This is a very
convenient basis for evaluating the loop corrections to the vertex,
since the $L$ and $T$ parts separate, as described in Appendix A.

The $\pm$ basis and the $LT$ basis are complementary in the
following sense. The $\pm$ basis is constructed out of combinations
of longitudinal ($+$) and transverse ($-$) momenta, so that for
example $(00+)=g_{\mu_1\mu_2}(p_{1\mu_3}+p_{2\mu_3}) = -
g_{\mu_1\mu_2}p_{3\mu_3}$ vanishes if the $\mu_3$ index is
contracted into a conserved current. Meanwhile, the $LT$ basis
distinguishes between parts of the vertex that do ($T$) and do not
($L$) vanish when dotted with longitudinal momenta. These
straightforward relations to current conservation and Ward
identities are essentially the reason these two bases are the most
convenient to work with.

 In our notation, the vertex can be written in the
 $LT$ basis as $\Gamma=\Gamma_L+\Gamma_T$, where
 \begin{eqnarray}\label{LTbasis}
 \Gamma_L &=& (\bar{A}_{12}\bar{a}_{12} + \bar{B}_{12}\bar{b}_{12}+\bar{C}_{12}\bar{c}_{12} + {\rm perms}) + \bar{S}\bar{s}
 \nonumber\\
 \Gamma_T &=& (\bar{F}_{12}\bar{f}_{12} + {\rm perms})
 +\bar{H} \bar{h},
 \end{eqnarray}
and the bar distinguishes this $LT$ basis from the $\pm$ basis
defined above in Eq.(\ref{pmbasis}). The relation between basis
tensors is given by
\begin{eqnarray}
 \bar{a}_{12} &=& 001-002 = \hat{a}_{12}\\
 \bar{b}_{12} &=& 001+002 = \hat{b}_{12}\nonumber\\
 \bar{c}_{12} &=& 211-212-  \pot (001-002) = {1\over 4}(\hat{c}_{12}-\hat{d}_{23}+\hat{d}_{31}-\hat{h}) - \pot \hat{a}_{12}\nonumber\\
 \bar{f}_{12} &=& \pot \Big(\ptt 001 -\pto 002\Big) -\Big(\ptt 211 -\pto 212\Big) \nonumber\\
              &=& {\pot \over 2} \Big(-p_3^2 \hat{a}_{12} + (p_1^2-p_2^2)\hat{b}_{12}\Big) \nonumber\\
  &+& {1\over 8} \Big(p_3^2(\hat{c}_{12}-\hat{d}_{23}+\hat{d}_{31}-\hat{h}) +(p_1^2-p_2^2)(\hat{d}_{12}-\hat{c}_{23}+\hat{c}_{31}-\hat{s})\Big)\nonumber\\
 \bar{h} &=& 231-312 -\Big( \pot (030-300) + \ptt (001-010) + \pto (200 - 002) \Big)\nonumber\\
 &=& {1\over 4}(\hat{h}+\hat{c}_{12}+\hat{c}_{23}+\hat{c}_{31}) + \half \Big(p_3^2 \hat{a}_{12} -(p_1^2-p_2^2) \hat{b}_{12}
 + p_1^2 \hat{a}_{23} - (p_2^2-p_3^2) \hat{b}_{23}
  \nonumber\\
 &+& p_2^2 \hat{a}_{31} -(p_3^2-p_1^2) \hat{b}_{31}\Big) \nonumber\\
 \bar{s} &=& 231+312 = {1\over
 4}(\hat{s}+\hat{d}_{12}+\hat{d}_{23}+\hat{d}_{31})\nonumber,
 \end{eqnarray}
and we used $\pot = (p_3^2-p_1^2-p_2^2)/2$.
 This implies the relation between form factors
\begin{eqnarray}\label{pmLTrel}
 A_{12} &=& \bar{A}_{12} - \pot \bar{C}_{12} -{p_3^2\over 2} \Big( \pot \bar{F}_{12} - \bar{H} \Big) \nonumber\\
 B_{12} &=& \bar{B}_{12} + {p_1^2-p_2^2\over 2} \Big( \pot \bar{F}_{12} - \bar{H} \Big) \\
 C_{12} &=& {1\over 4} \Big( \bar{H} +\bar{C}_{12} +{p_3^2\over2} \bar{F}_{12} +{p_1^2-p_3^2\over2} \bar{F}_{31}+{p_2^2-p_3^2\over2} \bar{F}_{23} \Big) \nonumber\\
 D_{12} &=& {1\over 4} \Big( \bar{S} +\bar{C}_{23}-\bar{C}_{31}  +{p_1^2\over2} \bar{F}_{23} - {p_2^2\over2} \bar{F}_{31}+{p_1^2-p_2^2\over2} \bar{F}_{12} \Big) \nonumber\\
 H &=&  {1\over 4} \Big( \bar{H} -\bar{C}_{12} -{p_3^2\over2} \bar{F}_{12} -\bar{C}_{23} -{p_1^2\over2} \bar{F}_{23}-\bar{C}_{31} -{p_2^2\over2} \bar{F}_{31}\Big) \nonumber\\
 S &=&  {1\over 4} \Big( \bar{S} + {p_2^2-p_1^2\over2} \bar{F}_{12} + {p_3^2-p_2^2\over2} \bar{F}_{23} + {p_1^2-p_3^2\over2} \bar{F}_{31}\Big)\nonumber,
 \end{eqnarray}
The unwritten form factors ($A_{23}$, etc.) and basis tensors can be
obtained trivially from the above equations by cyclic permutation
($g_{123}$). In doing so it is useful to keep in mind the properties
described under Eq.(\ref{3gvertex}), along with $\bar{F}_{ij} =
\bar{F}_{ji}$, $\bar{A}_{ij} = \bar{A}_{ji}$, $\bar{B}_{ij} =
-\bar{B}_{ji}$, $\bar{C}_{ij} = \bar{C}_{ji}$, while $\bar{H}$ and
$\bar{S}$ have the same transformation properties as $H$ and $S$,
respectively.

\section{Results for the Form Factors}

In this section, we will present the results for the form factors
in arbitrary dimension $d$ using dimensional regularization
(DREG), for gluons in the adjoint representation, and massless
quarks and scalars in arbitrary representations. The corrections
due to supersymmetric regularization and massive fermions,
scalars, and gauge bosons are given in detail in Appendices D and
E, respectively.

The gauge invariant vertex at one loop can be written as
 \begin{eqnarray}\label{fullvertex}
 &g& \!\!\!\!\! \Gamma_{\mu_1\mu_2\mu_3}^{abc}(p_1,p_2,p_3) =  gf^{abc}\Bigg[
 \Gamma_{\mu_1\mu_2\mu_3}^{(0)}(p_1,p_2,p_3)
 \nonumber\\
 &-& {ig^2\over 2}\Bigg( C_A G_{\mu_1\mu_2\mu_3}
 +2\sum_f T_f N_f Q_{\mu_1\mu_2\mu_3}+2\sum_s T_s N_s S_{\mu_1\mu_2\mu_3}
 \Bigg)\Bigg]
 \end{eqnarray}
 where the gluon ($G$), quark ($Q$), and scalar ($S$) integrals are
 \begin{eqnarray}\label{GQSdef}
 G_{\mu_1\mu_2\mu_3} &=& \int {d^dl\over (2\pi)^d} {1\over
 l_1^2l_2^2l_3^2} \Bigg(
 \Gamma_{\b\mu_1\g}^F(l_2,p_1,-l_3)\Gamma_{\g\mu_2\a}^F(l_3,p_2,-l_1)\Gamma_{\a\mu_3\b}^F(l_1,p_3,-l_2)
 \nonumber\\
 &+& 2(l_2+l_3)_{\mu_1}(l_3+l_1)_{\mu_2}(l_1+l_2)_{\mu_3}
 - 8l_1^2(g_{\mu_1\mu_2}p_{1\mu_3}-g_{\mu_1\mu_3}p_{1\mu_2})
 \nonumber\\
 &-&8l_2^2 (g_{\mu_2\mu_3}p_{2\mu_1}-g_{\mu_2\mu_1}p_{2\mu_3})
  - 8l_3^2 (g_{\mu_3\mu_1}p_{3\mu_2}-g_{\mu_3\mu_2}p_{3\mu_1})\Bigg)
\nonumber\\
 Q_{\mu_1\mu_2\mu_3} &=& \int {d^dl\over (2\pi)^d}
 {1\over l_1^2l_2^2l_3^2} {\rm
 Tr}[\g_{\mu_1}\slash{l}_3\g_{\mu_2}\slash{l}_1\g_{\mu_3}\slash{l}_2]
 \nonumber\\
 S_{\mu_1\mu_2\mu_3} &=& -\int {d^dl\over (2\pi)^d}
 {1\over
 l_1^2l_2^2l_3^2}(l_2+l_3)_{\mu_1}(l_3+l_1)_{\mu_2}(l_1+l_2)_{\mu_3}.
 \end{eqnarray}
 The gluon contribution was first derived in \cite{Cornwall:1989gv} using the pinch technique(PT),
 and is equivalent to the vertex obtained in the Background Field Method
 in quantum Feynman gauge(BFMFG).
 The quark and scalar integrals come straightforwardly from the one loop triangle
 diagrams.
 The notation and routing of the integral are defined in Fig.(\ref{fig:1}) such that
 $l_1=p_2+l_3$, $l_2=p_3+l_1$, $l_3=p_1+l_2$ and
 the tree level vertex $\Gamma^{(0)}$ and $\Gamma^F$ are defined as
 \begin{eqnarray}
 \Gamma_{\mu_1\mu_2\mu_3}^{(0)}(p_1,p_2,p_3) &=&
 g_{\mu_1\mu_2}(p_1-p_2)_{\mu_3}+g_{\mu_2\mu_3}(p_2-p_3)_{\mu_1}+g_{\mu_3\mu_1}(p_3-p_1)_{\mu_2}
 \nonumber\\
 \Gamma_{\b\mu_1\g}^F(l_2,p_1,-l_3) &=& 2p_{1\b}g_{\mu_1
 \g}-2p_{1\g}g_{\mu_1\b}- (l_2+l_3)_{\mu_1}g_{\b\g}.
 \end{eqnarray}

All massless integrals can be reduced to two basic scalar integrals:
\begin{eqnarray}\label{Jdef}
J \;&{\equiv}&\; J(p_1^2,p_2^2,p_3^2) =  \int {d^dl\over (2\pi)^d}
{1\over
 l_1^2l_2^2l_3^2} \nonumber\\
 J_1 \;&{\equiv}&\; J_1(p_1^2) =  \int {d^dl\over (2\pi)^d} {1\over
 l_2^2l_3^2} = \int {d^dl\over (2\pi)^d} {1\over
 l^2(l+p_1)^2},
\end{eqnarray}

These functions, and the massive integrals which are considered
later, are summarized in Appendices A and B, where $J$ is written in
terms of Clausen functions. In the following we will suppress the
momentum arguments and write our results in terms of
$J,J_1,J_2,J_3$.

\subsection{(Supersymmetric) relations between gluons, quarks, and scalars}

Before presenting the results for individual form factors, which are
somewhat lengthy, we will discuss the relationship between the
gluon(G), quark(Q), and scalar(S) contributions. We will end up
finding relations similar to those found in the context of
supersymmetric scattering amplitudes
\cite{Bern:1994zx}\cite{Bern:1994cg}\cite{Bern:1996je}. For a
generic form factor $F$, let us write the one-loop contribution as
 \bege\label{Fdef}
 F = ig^2\Big(C_AF_G + 2\sum_f T_f N_f F_Q + 2\sum_s T_s N_s F_S\Big),
 \ende
where the coupling constant $ig^2$ and group theory factors have
been pulled out. The standard notation is used, so that $C_A\equiv
C_2(G)=N_c$ for $SU(N_c)$, and ${\rm Tr} [t^a_f t^b_f] = T_f
\d^{ab}$. Thus, $F_Q$ stands for the contribution of one Dirac
fermion in the fundamental representation of $SU(N_c)$, or, due to a
symmetry factor of $\half$ for Weyl fermions, the contribution of
adjoint gluinos divided by $N_c$. Similarly, $F_S$ stands for the
contribution of one complex scalar in the fundamental
representation, or the contribution of a real scalar in the adjoint,
divided by $N_c$. These identifications will be used shortly.

After explicitly calculating the integrals in Eq.(\ref{GQSdef}), we
noticed that the gluon, quark, and scalar contributions have a
similar structure for each form factor. To make this explicit,
define the following sums for form factor $F$ :
 \begin{eqnarray}\label{sumdef}
 \Sigma_{QG}(F) &\equiv&  {(d-2)\over 2}F_Q + F_G
 \nonumber\\
 \Sigma_{SG}(F) &\equiv& (d-2)F_S - F_G.
 \end{eqnarray}

Although the results for each form factor are often long, these sums
are particularly simple, as can be seen in
Eqs.(\ref{QGsumLT},\ref{QGsumpm},\ref{QGsumsym}). For all form
factors in any basis, it also turns out that
 \bege\label{SGQG}
(d-10) \Sigma_{SG} = 8 \Sigma_{QG}
 \ende
 and $\Sigma_{QG}$ is always proportional to $d-10$.
The above two equations and the results of
Eqs.(\ref{QGsumLT},\ref{QGsumpm},\ref{QGsumsym}) can be used to
determine the $Q$ and $S$ contributions to any form factor, given
the gluon contributions written explicitly below. Furthermore,
Eqs.(\ref{sumdef},\ref{SGQG}) can be combined leading to
 \bege\label{GQSsum}
 F_G +4F_Q +(10-d)F_S = 0.
 \ende
Considering the very different origins of each form factor
(Eqs.(\ref{fullvertex},\ref{GQSdef})), it is remarkable that they
are related in such a simple manner. Note that no such analogous
relation holds for the gauge dependent vertex
\cite{Davydychev:1996pb}.

This type of relation hints at supersymmetry. To further understand
the content of these relations, we will consider various
supersymmetries in $d=4$.
 \begin{itemize}
 \item{\underline{\bf{N=1}} From the above definitions, it is clear that a vector
  superplet $V_1$ (gluons plus gluinos) contributes
  $ig^2N_c(F_G+F_Q)\equiv ig^2N_cF_{V_1}$ to a generic form factor $F$, while $N_{\Phi}$ chiral
  superplets contributes $ig^2N_{\Phi}(\half F_Q + F_S)\equiv ig^2N_{\Phi}F_{\Phi}$. By
  the sum rule Eq.(\ref{GQSsum}) in $d=4$, we have $F_{V_1}+6F_{\Phi}=0$.
  Thus any form factor can be written
  \bege
   F=ig^2(N_cF_{V_1}+N_{\Phi}F_\Phi)={ig^2\over3}\b_0^{(N=1)}F_{V_1},
  \ende
  where $\b_0^{(N=1)}= 3N_c-\half N_{\Phi}$ is the first $\b$ function
  coefficient. Hence the contributions of vector and chiral superplets have precisely the
  same functional form for each form factor. Furthermore, every form factor is proportional
  to $\b_0$ even though all but one of them are UV finite.
  }
 \item{\underline{\bf{N=2}} Here the vector superplet gives
 $ig^2N_c(F_G+2F_Q+2F_S)\equiv ig^2N_cF_{V_2}$, while $N_h$ hyperplets (a Weyl fermion of each helicity
 plus a doublet of complex scalars)
 yield $ig^2N_h(F_Q+2F_S)\equiv ig^2N_hF_h$. The sum rule Eq.(\ref{GQSsum})
 can be written as $F_{V_2}+2F_h=0$, and thus
 \bege
  F  = ig^2(N_cF_{V_2}+N_hF_h)={ig^2\over 2} \b_0^{(N=2)} F_{V_2},
 \ende
where $\b_0^{(N=2)} = 2N_c-N_h$.
 }
 \item{\underline{\bf{N=4}} Here the vector superplet (the only multiplet allowed) contributes
 $2 ig^2 N_c(F_G+4F_Q+6F_S)\equiv N_cF_{V_4}$, which is identically zero by
 the sum rule, which of course is a consequence of $\b_0^{(N=4)}=0$.
 }
 \end{itemize}

 Thus, the similarities between form factors in $d=4$
 are related to supersymmetric non-renormalization theorems.
 In particular, the exact conformal invariance
 of $N=4$ implies that the gauge-invariant three-gluon Green's function is not
 renormalized at any order in perturbation theory. Furthermore, at
 one-loop order there are not even finite corrections, as reflected in
Eq.(\ref{GQSsum}).

 Analogous results hold
 for supersymmetry in $d\neq 4$. Here we must be careful, because
 in the sum rules and form factors presented in this paper we worked in $d$
 dimensions everywhere except in the traces over gamma matrices,
 where we used the conventional rule of dimensional regularization
 ${\rm tr}[\g_\mu\g_\nu] = 4 g_{\mu\nu}$, and similarly for other traces.
 Properly working in integer valued $d$ dimensions, we should use
 ${\rm tr}[\g_\mu\g_\nu] = d_s(d) g_{\mu\nu}$, where the spinor dimension of
 the gamma matrices is
 \bege
 d_s(d) = \left( \begin{array}{ccc} 2^{d/2}  \\ 2^{(d-1)/2} \end{array} \right)\;\;{\rm for} \;\;
 \left( \begin{array}{ccc} d \;\;{\rm even} \\  d \;\;{\rm odd} \end{array} \right).
 \ende
Thus $F_{Q_d}\equiv {d_s(d)\over 4} F_Q$ is the contribution of a
Dirac fermion in $d$ dimensions, and we have
 \bege\label{GQSsumd}
 F_G+{16\over d_s(d)}F_{Q_d}+(10-d)F_S = 0
 \ende
 Rather than using Eq.(\ref{GQSsumd}), one can
alternatively use Eq.(\ref{GQSsum}) and be sure to count fermion
degrees of freedom in terms of $d=4$ spinors. Thus the Weyl fermions
of $d=6$ and the Weyl-Majorana fermions of $d=10$ are composed of
$2$ and $4$ Weyl fermions of four dimensions, respectively. From
this, it is straightforward to show that $d=6,N=2$ and $d=10,N=1$
gauge theory give vanishing contribution to every form factor. For
the $d=6,N=1$ case, one finds
 \bege
 F={ig^2\over 2} \beta_0 F_{V_1} \;\;\;\;\;\; \beta_0 =
 2N_c-N_{\Phi},
 \ende
where $\b_0$ is determined from Eq.(\ref{beta0}) in $d=6$.

Note that it is not straightforward to analytically continue
$d_s(d)$ into arbitrary non-integer $d$, which is the reason for the
simple dimensional regularization rule. However, the sum rule
expressed in Eq.(\ref{GQSsum}) {\it is} an analytic function of $d$
and thus represents an analytic continuation of supersymmetric
non-renormalization theorems to arbitrary $d$. This is intimately
related to the existence of a supersymmetry preserving regulator,
dimensional reduction (DRED), where vector degrees of freedom are
kept in four dimensions while the integrals are still performed in
$d$ dimensions. Around four dimensions, $d=4-2\e$, we have
$F_G+4F_Q+(6+2\e)F_S=0$ in dimensional regularization, and we see
that the $\e$ term plays the role of the so-called $\e$-ghosts of
DRED. We have calculated the form factors in DRED (see Appendix D)
and verified that
 \bege
 F_G+4F_Q+6F_S=0 \;\;\;\;\;\; {(\rm DRED)}.
 \ende
 In the preceding discussion of supersymmetries in various
 dimensions we implicitly used DRED.

The extension of these relations to the massive case is outlined in
Appendix E, where the full effects of internal massive fermions
(MQ), massive scalars (MS), and massive gauge bosons (MG) are
included, and the sum rule becomes
 \bege
 F_{MG}+4F_{MQ}+(9-d)F_{MS}=0
 \ende
 in DREG while in DRED the only change is $9-d$ is replaced by $5$.
Note that the external gluons remain massless and unbroken, so the
internal massive gauge bosons might be the colored heavy gauge
bosons arising in GUT models. The change of $10-d$ in the massless
case to $9-d$ in the massive case reflects the fact that massive
gauge bosons ``eat" one scalar degree of freedom.

 It should be emphasized that relations such as Eq.(\ref{GQSsum}) do not exist
 for the gauge-dependent three-gluon vertex \cite{Davydychev:1996pb}, since the
 gluon contributions depend on the gauge-parameter, while the
 quarks and scalars do not. Indeed, it is
 uniquely the pinch technique (or equivalently BFM in quantum Feynman gauge)
 Green's function which satisfies this homogeneous sum rule.
 For example, calculating in the BFM with $\xi_Q \neq 1$, leads to
 a nonzero RHS of Eq.(\ref{GQSsum}).

 Since the sum rule applies to all form factors, one finds
 \bege
 G_{\mu_1\mu_2\mu_3} +
 4Q_{\mu_1\mu_2\mu_3}+(10-d)S_{\mu_1\mu_2\mu_3}=0,
 \ende
 which is remarkable given the expressions in Eq.(\ref{GQSdef}). One
 can explicitly show this by performing the trace in
 Eq.(\ref{GQSdef}) and some tedious algebra to rearrange
 the $\Gamma^F\Gamma^F\Gamma^F$ term. This can also be
 seen in the second order formalism of the
 BFM  \cite{Bern:1994zx}\cite{Bern:1994cg}\cite{Bern:1996je}.

 A similar relation holds for the one-loop gauge-invariant (pinch-technique)
 gluon two-point function in $d$ dimensions,
 \bege
 \Pi_{\mu_1\mu_2}^{ab}(p) = \d^{ab}(p^2g_{\mu_1\mu_2}-p_{\mu_1}p_{\mu_2})\Pi(p^2)
 \;{\equiv}\;ig^2\d^{ab} \Big( N_c G_{\mu_1\mu_2}
 +2\sum_f T_f N_f Q_{\mu_1\mu_2}+2\sum_s T_s N_s S_{\mu_1\mu_2}
 \Big),
 \ende
where from Eqs.(\ref{Pi},\ref{beta0}) below we find
 \bege
 G_{\mu_1\mu_2} +
 4Q_{\mu_1\mu_2}+(10-d)S_{\mu_1\mu_2}=0.
 \ende

Unfortunately, analogous relations do not hold for higher
gauge-invariant gluon $n$-point functions. This is essentially
because the color and spacetime indices mix nontrivially. However,
inhomogeneous relations of the form
 \bege
 G+4Q+(10-d)S = {\rm simple}
 \ende
still hold \cite{Bern:1994zx}\cite{Bern:1994cg}\cite{Bern:1996je},
where ``simple" means an integral with fewer powers of loop momenta
in the numerator. In the four-gluon case, this is just a simple
scalar integral with no powers of loop momentum in the numerator.
These loop momentum counting rules have been derived in the second
order formalism, which is reviewed in \cite{Bern:1996je}. Note that
the Ward ID for the four-gluon vertex \cite{Papavassiliou:1992ia}
relates it to the three-gluon vertex, and thus the longitudinal
parts of the four-gluon vertex must satisfy the homogeneous sum
rules $F_G+4F_Q+(10-d)F_S=0$.

It is interesting to see if extensions of these sum rules apply to
two-loop calculations, where the supersymmetric Yukawa vertices must
be taken into account. As a first application, the two-loop pinch
technique gluon self-energy has been calculated including finite
terms. Interestingly, the finite terms do not vanish for $N=4$ SUSY,
so it appears that the homogeneous sum rule in Eq.(\ref{GQSsum})
does not have a counterpart at two loops. In any case, the finite
parts of the two loop result allow for an improved extraction of the
PT couplings from data as well as giving the three loop beta
function. This calculation will be reported elsewhere
\cite{2loopPTSE}.

Now explicit expressions for the form factors will be given, first
in the $LT$ basis, and then in the $\pm$ basis.

\subsection{The longitudinal form factors}

It is straightforward to solve the Ward identity
(Eq.(\ref{wardid2})) for the ten longitudinal form factors, defined
in Eq.(\ref{LTbasis}), in terms of the gluon self-energy function
$\Pi$ defined by $\Pi_{\mu\nu}^{ab}(p) = \d^{ab}(p^2g_{\mu\nu} -
p_{\mu}p_{\nu})\Pi(p^2)$. Note that this is not the usual
gauge-dependent self-energy, but rather the gauge-invariant pinch
technique self-energy. At one loop in $d$ dimensions this is given
by
 \bege\label{Pi}
 \Pi(p^2) = ig^2\b_0(d) \int {d^dl\over (2\pi)^d} {1\over
 l^2(l+p)^2},
 \ende
 where $\beta_0(d)$ is given by
 \begin{eqnarray}\label{beta0}
 \beta_0(d) &=& {7d-6\over
 2(d-1)}C_2(G)-{2(d-2)\over(d-1)}\sum_f T_fN_f-{1\over(d-1)}\sum_sT_sN_s
 \end{eqnarray}
 for massless gluons, quarks, and complex scalars. The mass-dependent
 results are given in section 5 and Appendix E.
 This result holds for dimensional regularization (DREG), whereas for dimensional reduction
 (DRED) the gluon coefficient changes from $(7d-6)$ to $(8d-10)$.

 The longitudinal form factors are given by
 \begin{eqnarray}\label{LFFs}
 \bar{A}_{12} &=& {\Pi(p_1^2)+\Pi(p_2^2)\over 2} \nonumber\\
 \bar{B}_{12} &=& {\Pi(p_1^2)-\Pi(p_2^2)\over 2} \\
 \bar{C}_{12} &=&  {\Pi(p_1^2)-\Pi(p_2^2)\over p_1^2-p_2^2} \nonumber\\
 \bar{S}&=&0\nonumber,
 \end{eqnarray}
and of course cyclic permutations yield results for $\bar{A}_{23}$,
etc. Note that one of the 14 form factors vanishes to all orders and
only the $\bar{A}$ form factors contain UV divergences.

In the notation of Eqs.(\ref{Jdef},\ref{Fdef}) we have
 \begin{eqnarray}
  \bar{A}_{12}(G) &=& {7d-6\over 2(d-1)}\half (J_1+J_2)\nonumber\\
  \bar{A}_{12}(Q) &=& {2-d\over(d-1)}\half (J_1+J_2)\\
  \bar{A}_{12}(S) &=& -{1\over 2(d-1)}\half (J_1+J_2)\nonumber,
 \end{eqnarray}
 and similarly for the $\bar{B}$ and $\bar{C}$ form factors.

\subsection{The transverse form factors}

These form factors cannot be determined from the Ward ID, and must
be calculated explicitly. The algorithm used is briefly described in
Appendix A.

Due to the lengthy expressions, the following shorthand notation
will be used for the kinematic invariants:
\begin{eqnarray}
 a = p_1^2 \;\;\;\;\; b=p_2^2 \;\;\;\;\; c=p_3^2 \;\;\;\;\;
 \a = p_1 \!\cdot\! p_2 \;\;\;\;\; \b = p_2 \!\cdot\! p_3 \;\;\;\;\; \g = p_3 \!\cdot\! p_1
 \end{eqnarray}
 We also define the symmetric invariants
 \begin{eqnarray}
  \CQ &=& \a+\b+\g \nonumber\\
  \CK &=& \a\b+\b\g+\g\a \\
  \CP &=& \a\b\g\nonumber.
 \end{eqnarray}
Note that the dot products can be written in terms of the
virtualities $\a = (c-a-b)/2$, $\b = (a-b-c)/2$, $\g = (b-c-a)/2$,
or vice versa $a=-\a-\g$, $b=-\a-\b$, $c=-\b-\g$, but the formulae
are simpler and more transparent when selectively written in terms
of both $\a,\b,\g$ and $a,b,c$.

 We will only write the full gluon contribution explicitly, since the quark and scalar
 parts can be determined from the results of section 3.1
 (see Eqs.(\ref{sumdef},\ref{SGQG},\ref{GQSsum})) and the
 quark-gluon sum rules for the transverse form factors, which are
 \begin{eqnarray}\label{QGsumLT}
 \Sigma_{QG}(\bar{F}_{12}) &=& -{(d-10) \over2\CK} \Bigg( \a J +
 {2\a(J_1-J_2) -\b(J_2-J_3)-\g(J_3-J_1)\over \b-\g}\Bigg)
 \nonumber\\
 \Sigma_{QG}(\bar{H}) &=& {(d-10) \over2} J.
 \end{eqnarray}

The gluon contributions to the transverse form factors in the $LT$
basis are
 \begin{eqnarray}
  \bar{F}_{12}(G) &=& {1\over 2\CK^2}\Bigg( J\Big( 10\CP
  +c(\CK-7\a^2-3\b\g)\Big) \\
  &+&\Big( 1-{(d+1)\b\g\over \CK} \Big) \Big( \CP J+\a\g J_1 +\a\b J_2 + \b\g J_3 -{\CK\over d-1}(J_1+J_2+J_3) \Big) \nonumber\\
 &+&{7d-6\over 2(d-1)(d-2)} \Big[ 8\CP +(d-4)\a c^2+(d-2)(4\CK \a-c\b\g) \Big]{J_1-J_2\over a-b}\nonumber\\
 &+& {1\over d-1} \Big[2\b^2(d-2)-{5d-2\over 2}\CK -{7d-6\over d-2}\a\b-{2d^2-15d+14\over d-2}\a\g \Big](J_2-J_3)\nonumber\\
 &-& {1\over d-1} \Big[2\g^2(d-2)-{5d-2\over 2}\CK -{7d-6\over d-2}\a\g-{2d^2-15d+14\over d-2}\a\b \Big](J_3-J_1)
 \Bigg)\nonumber
 \end{eqnarray}
and
 \begin{eqnarray}
  \bar{H}(G) &=& -{1\over 2\CK^2}\Bigg( J\Big[ 8\CK^2 +(d-2)\CP\CQ +(d+1){abc\CP\over
  \CK}\Big]\nonumber\\
  &+&{d-2\over d-1}\Big[ \a(\CK-2\a\g)(J_1-J_2) + \b(\CK-2\b\a)(J_2-J_3) +
  \g(\CK-2\b\g)(J_3-J_1)\Big]
  \nonumber\\
  &+&\CP {d+1\over d-1}\Big[ -(J_1+J_2+J_3) +{d-1\over \CK}(\a\g J_1+\a\b J_2 + \b\g J_3)\Big]
  \Bigg).
 \end{eqnarray}

\subsection{The Form Factors in the Physical Basis}

Now we will present the results in the physical $\pm$ basis
(Eq.(\ref{pmbasis})), before symmetrization, Eq.(\ref{sympmdef}),
since this is the most convenient way to present the results. Of
course, these results can be obtained from the relation between the
$\pm$ basis and the $LT$ basis given in Eq.(\ref{pmLTrel}), but we
write them explicitly for future convenience and phenomenological
applications.

The quark-gluon sums are given by

\begin{eqnarray}\label{QGsumpm}
 \Sigma_{QG}(A_{12})&=&{(d-10)\over 4\CK}\Big( abc J +a\b J_1 + b\g J_2 + c\a J_3  \Big)  \nonumber\\
 \Sigma_{QG}(B_{12})&=&{(d-10)\over 4\CK}\Big( (\g-\b)a b J +(2\a+\b)aJ_1-(2\a+\g)bJ_2-\a(\b-\g)J_3\Big) \nonumber\\
 \Sigma_{QG}(C_{12})&=& -{(d-10)\over 4\CK}\Big( \a c J +\g J_1 +\b J_2 +cJ_3 \Big) \nonumber\\
 \Sigma_{QG}(D_{12})&=& 0 \\
 \Sigma_{QG}(H)&=& 0  \nonumber\\
 \Sigma_{QG}(S)&=& 0  \nonumber,
 \end{eqnarray}
and the remaining sums (for $A_{23}$, etc.) are related trivially by
permutations
 $\a\raw \b \raw \g \raw \a$, $ a\raw b\raw c\raw a$, $J_1\raw
 J_2\raw J_3\raw J_1$.

The gluon form factors in $d$ dimensions are
 \begin{eqnarray}
 -4\!\!\!\!&\CK^2&\!\!\!\! A_{12}(G)= abc J(7\CK +\b\g) +a J_1\Big( 7\CK \b +\b^2\g +\CK \g{d-2\over d-1}\Big) \\
 &+& b J_2\Big( 7\CK \g +\b\g^2 +\CK \b{d-2\over d-1}\Big) +c J_3\Big( 7\CK \a + \CP +\CK c{d-2\over d-1}\Big)
 \nonumber\\
 -4\!\!\!\!&\CK^2&\!\!\!\! B_{12}(G)= {ab J}(7\CK +\b\g)(\g-\b)
  + a J_1\Big( 7\CK \b -b\g(\b-\g) +\CK {2\a(7d-6)+\g\over d-1}\Big) \nonumber\\
 &-& b J_2\Big( 7\CK \g +a\b(\b-\g) +\CK {2\a(7d-6)+\b\over d-1}\Big)
 + (\g-\b)J_3 \Big( 7\CK \a + \CP +\CK c{d-2\over d-1}\Big)\nonumber
 \end{eqnarray}
 \begin{eqnarray}
  16\!\!\!\!&\CK^3&\!\!\!\! C_{12}(G)= c J\Bigg( 3\CK^2(10\a+c)+\CK(\a^3-6c\b\g)
  +\CP\CK(d+4)-\CP(d+1)(\a^2+2\b\g) \Bigg)\nonumber\\
 &+& {J_1}\Bigg( \CK^2\Big[{(3-2d)\over d-1}\a -\g{(d^2-30d+24)\over d-1}+3\b\Big]
 +\CP(d+1)\Big[{\CK\over d-1} + \g(2\CQ-3\a)\Big] \nonumber\\
 &+&\g\CK\Big[{(d^2-3)\over d-1}\a^2-6\b^2+{(d^2-8d+9)\over d-1}\b\g-4\g^2{(d-2)\over d-1}\Big]\Bigg)\\
&+& {J_2}\Bigg( \CK^2\Big[{(3-2d)\over d-1}\a -\b{(d^2-30d+24)\over d-1}+3\g\Big]+\CP(d+1)\Big[{\CK\over d-1} + \b(2\CQ-3\a)\Big] \nonumber\\
 &+&\b\CK\Big[{(d^2-3)\over d-1}\a^2-6\g^2+{(d^2-8d+9)\over d-1}\b\g-4\b^2{(d-2)\over d-1}\Big]\Bigg)\nonumber\\
 &+&{cJ_3}\Bigg( {(30d-31)\over d-1}\CK^2+ \CK\Big[\a^2-4c^2{(d-2)\over d-1}-{(d^2-4d+1)\over d-1}\b\g\Big]
 +(d+1)\CP(2\CQ-3\a)\Bigg)\nonumber
\end{eqnarray}
\begin{eqnarray}
 16\!\!\!\!&\CK^3&\!\!\!\! D_{12}(G)=ab(a-b)J\Big( \CK(\CQ+2\a)-(d+1)\CP
 \Big) - {aJ_1} \Bigg( {d^2-4\over d-1}\CK^2 \\
 &+&\CK\Big[\b^2-\b\g{(d^2-3)\over d-1}-\a\b{(d^2-4d+1)\over d-1}+4\a^2{(d-2)\over d-1}\Big] -(2\a+\b)\CP(d+1)  \Bigg)\nonumber\\
 &+& {bJ_2} \Bigg( {(d^2-4)\over d-1}\CK^2
 +\CK\Big[\g^2-\b\g{(d^2-3)\over d-1}-\a\g{(d^2-4d+1)\over d-1}+4\a^2{(d-2)\over d-1}\Big]\nonumber\\
 &-&(2\a+\g)\CP(d+1)  \Bigg) + {(a-b)J_3} \Bigg( {(4d-7)\over d-1}\CK^2
 +\CK\Big[3\a^2-\b\g{(d^2-3)\over d-1}\Big]-\a\CP(d+1)\Bigg)\nonumber
 \end{eqnarray}
 \begin{eqnarray}
 16\!\!\!\!&\CK^3&\!\!\!\!H(G) = abcJ \Big(\CP(d+1)-\CK\CQ\Big)\nonumber\\
 &+& aJ_1\Bigg( {3-2d\over d-1}\CK^2 +\Big[ {d^2-3\over d-1}\a\g -\b^2\Big]\CK
 +(d+1)\b\CP\Bigg)\nonumber\\
 &+& bJ_2\Bigg( {3-2d\over d-1}\CK^2 +\Big[ {d^2-3\over d-1}\a\b-\g^2\Big]\CK
 +(d+1)\g\CP\Bigg)\\
 &+& cJ_3\Bigg( {3-2d\over d-1}\CK^2 +\Big[ {d^2-3\over d-1}\b\g-\a^2\Big]\CK
 +(d+1)\a\CP\Bigg)\nonumber
 \end{eqnarray}
 \begin{eqnarray}
 16\!\!\!\!&\CK^3&\!\!\!\!S(G) =
 (a-b)(b-c)(c-a)J\Big(3\CK\CQ-(d+1)\CP\Big)\\
 &+& (b-c)J_1 \Bigg( {3\CK^2\over d-1} +\CK\Big[ 4a^2{d-2\over
 d-1}+\a\g{d^2-4d+1\over d-1}-3\b^2\Big]-(d+1)\CP(2\CQ-3\b)\Bigg)\nonumber\\
 &+& (c-a)J_2 \Bigg( {3\CK^2\over d-1} +\CK\Big[ 4b^2{d-2\over
 d-1}+\a\b{d^2-4d+1\over d-1}-3\g^2\Big]-(d+1)\CP(2\CQ-3\g)\Bigg)\nonumber\\
 &+& (a-b)J_3 \Bigg( {3\CK^2\over d-1} +\CK\Big[ 4c^2{d-2\over
 d-1}+\b\g{d^2-4d+1\over d-1}-3\a^2\Big]-(d+1)\CP(2\CQ-3\a)\Bigg)\nonumber
 \end{eqnarray}

Now we turn to the physical symmetrized basis. From
Eq.(\ref{sympmdef}), we see that for any triplet of form factors,
say $A_{ij}$, we have
 \begin{eqnarray}
 A_0 &=& {1\over 3} \Big( A_{12}+A_{23}+A_{31}\Big)\nonumber\\
 A_+ &=& {1\over 3} \Big( A_{12}+\l A_{23}+ \lb A_{31}\Big)\;\equiv\; A_1+iA_2\\
 A_- &=& {1\over 3} \Big( A_{12}+\lb A_{23}+ \l A_{31}\Big)\;\equiv\; A_1-iA_2\nonumber,
 \end{eqnarray}
 where we have defined
 \begin{eqnarray}
 A_1&=&{1\over 3} \Big( A_{12}-\half (A_{23}+A_{31})\Big)\nonumber\\
 A_2&=&{\sqrt{3}\over 6} \Big( A_{23}-A_{31} \Big).
 \end{eqnarray}

 $A_1$ and $A_2$ correspond to the real and imaginary
 parts of $A_{\pm}$ only when $J,J_1,J_2,J_3$ are real.
 This occurs (in the massless case) when $\CK>0$, which can only happen
 if all three gluon virtualities are of the same sign, either all spacelike or
 all timelike. This is often not the case for real problems.
 In general, however, it can be shown that
 \begin{eqnarray}
 A_{\pm}^*(a,b,c) &=& A_{\mp}(-a,-b,-c)\nonumber\\
 B_0^*(a,b,c) &=& B_0(-a,-b,-c)\nonumber\\
 B_{\pm}^*(a,b,c) &=& B_{\mp}(-a,-b,-c)\nonumber\\
 C_0^*(a,b,c) &=& -C_0(-a,-b,-c)\nonumber\\
 C_{\pm}^*(a,b,c) &=& -C_{\mp}(-a,-b,-c)\\
 D_0^*(a,b,c) &=& -D_0(-a,-b,-c)\nonumber\\
 D_{\pm}^*(a,b,c) &=& -D_{\mp}(-a,-b,-c)\nonumber\\
 H^*(a,b,c) &=& -H(-a,-b,-c)\nonumber\\
 S^*(a,b,c) &=& -S(-a,-b,-c)\nonumber.
 \end{eqnarray}

Furthermore, all of the above form factors except for $A_0$ are
scale invariant,
 \bege
 F(\l a,\l b,\l c) = F(a,b,c) = F(a/c,b/c,1)\;\;\;\;\;\; \l>0.
 \ende

Only $A_0$ is not scale invariant and does not satisfy a simple
reality condition.

The quark-gluon sums are given by
 \begin{eqnarray}\label{QGsumsym}
 \Sigma_{QG}(A_0)&=& {(d-10)\over 4\CK}\Phi_0 \;\;\;\;\; \Sigma_{QG}(A_{\pm})=0  \nonumber\\
 \Sigma_{QG}(B_0)&=& -{(d-10)\over 8\CK}B_0(G) \nonumber\\
 \Sigma_{QG}(B_{\pm})&=& {(d-10)\over 36\CK}\Bigg( -3\Big(\CQ \Phi_2 + \CK J(\b-\g)\Big)
 {\pm} i\sqrt{3}\Big( \CQ \Phi_1 - \CK J(\CQ-3\a)\Big) \Bigg)\nonumber\\
 \Sigma_{QG}(C_0)&=& {(d-10)\over 6} J \nonumber\\
 \Sigma_{QG}(C_{\pm})&=& {(d-10)\over 24\CK}\Bigg( \Phi_1 {\pm}i\sqrt{3} \Phi_2 \Bigg) \\
 \Sigma_{QG}(D_0)&=&  \Sigma_{QG}(D_{\pm}) = 0 \nonumber\\
 \Sigma_{QG}(H)&=& \Sigma_{QG}(S)=0\nonumber
 \end{eqnarray}
where we have defined the commonly occuring functions
 \begin{eqnarray}
 \Phi_0 &=& abcJ+a\b J_1+b\g J_2 +c\a J_3 \nonumber\\
 \Phi_1 &=& (\CK-3\b\g)J-3\g J_1-3\b J_2+3(\b+\g)J_3\\
 \Phi_2 &=& \a(\b-\g)J+(2\a+\g)J_1-(2\a+\b)J_2+(\b-\g)J_3\nonumber.
 \end{eqnarray}

 From the definition of $\Sigma_{QG}$ in Eq.(\ref{sumdef})
 we see that the quark and gluon (and thus scalar, by Eq.(\ref{SGQG}))
contributions have the same functional form for the seven form
factors which have a zero in the above. Letting $F$ stand for
$A_{\pm},D_0,D_{\pm},H$ or $S$, we find that
 \bege
 F = ig^2 \Big( N_c- {4\over d-2} \sum_f T_f N_f + {2\over d-2} \sum_s T_s N_s \Big)
 F(G),
 \ende
which, in $d=4$ QCD, reduces to $F=ig^2(N_c-N_f)F(G)$.

 In addition, both $A_0$ and $B_0$ are
governed by one function, since they satisfy different sum rules. In
particular, the tree level tensor structure has coefficient
 \bege
 A_0 = -{ig^2\over 2\CK}\left( {11\over 3}C_A-{2(3d-8)\over 3(d-2)}\sum_f T_fN_f-{2\over 3(d-2)}\sum_sT_sN_s
 \right)\Phi_0.
 \ende
This form factor will be discussed in more detail in the next
section.

Also, one finds from explicit calculation that the scalar
contribution to $B_0$ vanishes,
 \bege
 B_0(S) = 0 \;\;\;\;\;\;\;\; B_0(G)+4B_0(Q)=0,
 \ende
and thus
 \begin{eqnarray}
 B_0 &=& ig^2\left(N_c - \half \sum_f T_f N_f\right)B_0(G)\;\;\;\;\;{\rm where} \nonumber\\
 B_0(G) &=& -{2\over 3\CK} \Bigg( (a-b)(b-c)(c-a)J + (b-c)(2\CQ-3\b)J_1 + (c-a)(2\CQ-3\g)J_2\nonumber\\
 &+& (a-b)(2\CQ-3\a)J_3 \Bigg)
 \end{eqnarray}

 Finally, since $\bar{S}=0$ exactly, we know
that our fourteen dimensional basis is degenerate, which is
reflected in the fact that $S+3D_0=0$. Hence we define a new basis
tensor $\hat{d}_0'=\hat{d}_0-3\hat{s}$ so that
 $D_0\hat{d}_0+S\hat{s} = D_0(\hat{d}_0-3\hat{s}) = D_0\hat{d}_0'$.

Thus, we find that eight of the thirteen nonzero form factors have
the same functional form for gluons, quarks, and scalars. Only the
five form factors $B_{\pm},C_0$ and $C_{\pm}$ do not. These
statements are basis dependent. One can always find bases where none
of the form factors have a vanishing $QG$ sum rule. In the course of
our calculations, we found that the $\pm$ basis gives the maximum
number of such zeroes among bases which are reasonable and contain
the tree-level tensor structure. In this sense the (symmetric) $\pm$
basis is the simplest and most compelling. We will see in the next
section that this is also the most convenient basis for perturbative
calculations. Of course, as discussed in a previous section, with
supersymmetry every form factor is proportional to $\b_0$, and so
supermultiplets are governed by the same function in any basis.

\section{Three-Gluon Vertex in Perturbation Theory}

Applying the pinch-technique (PT) construction to the three-gluon
vertex occurring in a physical process involving three external
on-shell legs, one arrives at a dressed tree-level skeleton graph,
dressed with pinch-technique vertices and self-energies as shown in
Fig. 2.
\begin{figure}[htb]
 \centering \includegraphics[height=2.5in]{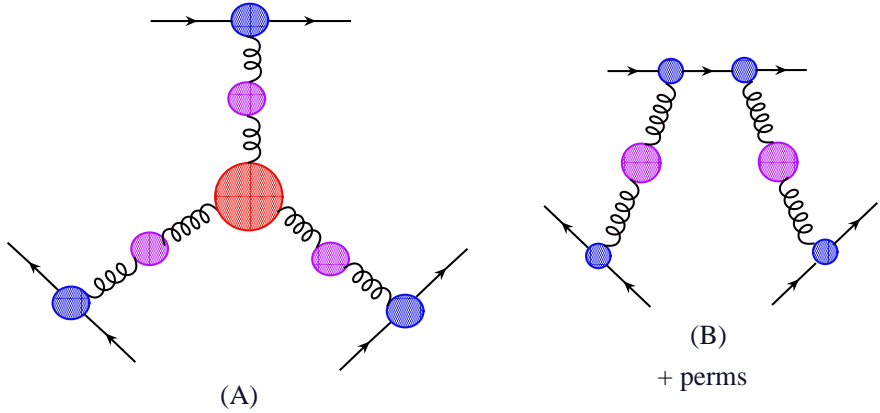}
 \caption[*]{The tree-level skeleton graphs dressed with pinch-technique vertices
 and self-energies, which are used to define effective charges both for the
 quark-quark-gluon vertex (depending on a single gluon virtuality) and for the three-gluon vertex,
 which depends on three different momenta.}
 \label{fig:2}
\end{figure}
Generically the amplitude of the three-gluon graph can be written
 \bege\label{Mamp}
 {\cal M} = {\cal C} g_0^4 V_1^{\nu_1}V_2^{\nu_2}V_3^{\nu_3}
 D_{\mu_1\nu_1}(k_1)D_{\mu_2\nu_2}(k_2)D_{\mu_3\nu_3}(k_3)\Gamma_{\mu_1\mu_2\mu_3}(k_1,k_2,k_3),
 \ende
where ${\cal C}$ is the overall color factor and $g_0$ is the bare
coupling. The PT vertices $V_i$ are for gluons coupled to external
particles, whose indices are suppressed; this is shown in Fig. 2 for
external quarks. $\Gamma = \sum_{i=1}^{13} F_i\hat{f}_i$ is the
gauge invariant three gluon vertex, whose thirteen form factors
($F_i$) are given in the preceding section. Finally the ``gauge
invariant" PT gluon propagator is
 \begin{eqnarray}\label{prop}
 D_{\mu\nu}(k) &=& {1\over k^2} \left( {t_{\mu\nu}(k)\over 1+\Pi(k^2)} + \xi l_{\mu\nu}(k) \right)\nonumber\\
 t_{\mu\nu}(k) &=& \left\{ \begin{array}{ccc} g_{\mu\nu}-{k_\mu k_\nu\over k^2}
  \\ g_{\mu\nu} -{n_\mu k_\nu + k_\mu n_\nu \over n\cdot k} \end{array}
  \right\}
  {\rm in} \left\{ \begin{array}{ccc} {\rm covariant} \\ {\rm axial} \end{array} \right\} {\rm gauges} \nonumber\\
 l_{\mu\nu}(k) &=& \left\{ \begin{array}{ccc} {k_\mu k_\nu\over k^2}
  \\ {k_\mu k_\nu\over (n\cdot k)^2} \end{array} \right\}
  {\rm in} \left\{ \begin{array}{ccc} {\rm covariant} \\ {\rm axial} \end{array} \right\} {\rm gauges},
 \end{eqnarray}
where $\xi$ is the gauge fixing parameter. This is ``gauge
invariant" in the maximal sense, i.e. the gauge dependence comes
only from the tree level terms, and in particular $\Pi(k^2)$ is
totally gauge invariant.

Regardless of whether the external particles are quarks, gluons, or
scalars, the vertices satisfy $k_1^{\mu}V_{1,\mu} = 0$ when these
particles are on shell (OS). One can then show that the gauge
dependent terms coming from Eq.(\ref{prop}) vanish in the full
amplitude consisting of all of the graphs in Fig.(\ref{fig:2}). This
can be seen trivially in the covariant gauges where the gauge
cancelations occur graph by graph, and with some work in axial
gauges, where the cancelation occurs between all of the graphs. In
the latter case, one must use the fact that the three gluon vertex
satisfies the Ward ID in Eq.(\ref{wardid}). Therefore we can take
 \bege
 D_{\mu\nu}(k) \raw {g_{\mu\nu}\over k^2(1+\Pi(k^2))}.
 \ende
Also, in the $\pm$ basis (Eq.(\ref{pmbasis})) any tensor with a
$'+'$ in any slot gives vanishing contribution to ${\cal M}$. For
example,
$(+00)=(k_2+k_3)_{\mu_1}g_{\mu_2\mu_3}=-k_{1,\mu_1}g_{\mu_2\mu_3}$,
and $k_{1,\mu_1}$ dots into $V_{1,\mu_1}$ yielding zero. Hence only
$(00-),(-00),(0\!-\!0)$, and $\mmm$ contribute, and we find
 \begin{eqnarray}
{\cal M} = {{\cal C} g_0^3 \over
(1+\Pi(k_1^2))(1+\Pi(k_2^2))(1+\Pi(k_3^2))}
{V_1^{\mu_1}V_2^{\mu_2}V_3^{\mu_3}\over k_1^2 k_2^2 k_3^2}
 g_0 \Gamma_{\mu_1\mu_2\mu_3}^{OS}(k_1,k_2,k_3),
 \end{eqnarray}
 where the three-gluon vertex connected to on-shell (OS) external
 particles is
 \begin{eqnarray}\label{GammaOS}
 \Gamma_{\mu_1\mu_2\mu_3}^{OS}(k_1,k_2,k_3) = (1+A_0) \hat{a}_0 + A_+ \hat{a}_+ + A_- \hat{a}_- + H
 \hat{h},
 \end{eqnarray}
in the notation of section 2 where hatted objects are three index
basis tensors.
\begin{figure}[htb]
 \centering \includegraphics[height=2.5in]{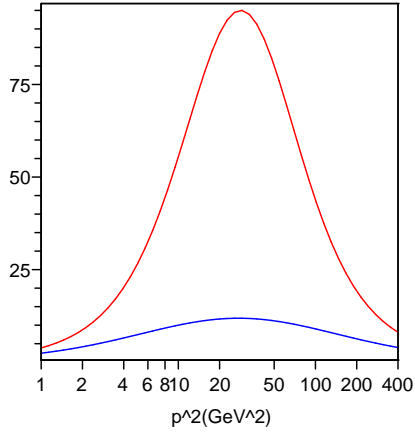} \caption[*]{The effective scale
 $Q^2_{eff}(10 \;{\rm GeV}^2,10 \;{\rm GeV}^2,p^2)$ is the lower blue curve, while
 $Q^2_{eff}(-10 \;{\rm GeV}^2,-10 \;{\rm GeV}^2,p^2) = Q^2_{eff}(10 \;{\rm
 GeV}^2,10 \;{\rm GeV}^2,-p^2)$ is the upper red curve. These both
 asymptote to zero, although very slowly for the upper curve.
 } \label{fig:3}
\end{figure}

\begin{figure}[htb]
 \centering \includegraphics[height=2.5in]{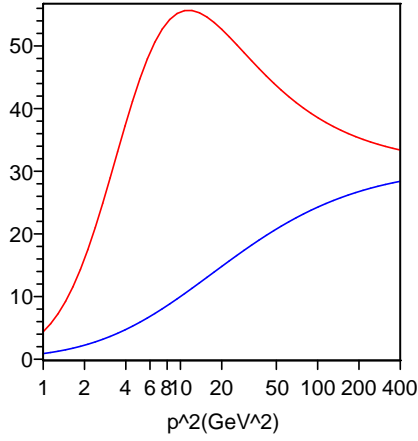} \caption[*]{The effective scale
 $Q^2_{eff}(10 \;{\rm GeV}^2,p^2,p^2)$ is the lower blue curve, while
 $Q^2_{eff}(-10 \;{\rm GeV}^2,p^2,p^2) = Q^2_{eff}(10 \;{\rm
 GeV}^2,-p^2,-p^2)$ is the upper red curve. These both asymptote to
 $10e^\Omega\;{\rm GeV^2} \approx 31.25 \;{\rm GeV^2}$.
 } \label{fig:4}
\end{figure}

Now one naturally defines PT effective charges by\footnote{
Eq.(\ref{qqgptdef}) holds for external fermions or scalars, but for
gluons one would instead have three additional three gluon effective
couplings, as is clear from the derivation of Eq.(\ref{3geff}).}
 \bege\label{qqgptdef}
 g^2(k_i^2)\equiv {g_0^2\over 1+\Pi(k_i^2)} \;\;\;\;
 i=1,2,3.
 \ende
 Since we only
have a single power of $g_0$ for each $1/(1+\Pi(k_i^2))$, this
leaves $\prod_{i=1}^3{1/\sqrt{1+\Pi(k_i^2)}}\approx
1-\half(\Pi(k_1^2)+\Pi(k_2^2)+\Pi(k_3^2))$ to be absorbed into the
three gluon vertex. Thus we have
 \begin{eqnarray}
 \prod_{i=1}^3{1\over \sqrt{1+\Pi(k_i^2)}}
 \Gamma_{\mu_1\mu_2\mu_3}^{OS}(k_1,k_2,k_3) &=&
 (1+\tilde{A}_0) \hat{a}_0 + A_+ \hat{a}_+ + A_- \hat{a}_- + H\hat{h} \nonumber\\
\tilde{A}_0(k_1^2,k_2^2,k_3^2) \equiv A_0(k_1^2,k_2^2,k_3^2)
\!\!\!&-&\!\!\! \half \big(\Pi(k_1^2)+\Pi(k_2^2)+\Pi(k_3^2)\big)
\end{eqnarray}
and
 \bege
 {\cal M} = {\cal C} g(k_1^2)g(k_2^2)g(k_3^2) {V_1^{\mu_1}V_2^{\mu_2}V_3^{\mu_3}\over k_1^2 k_2^2 k_3^2}
 g_0 \Big[(1+\tilde{A}_0) \hat{a}_0 + A_+ \hat{a}_+ + A_- \hat{a}_- + H
 \hat{h}\Big].
 \ende
This naturally leads to the effective coupling of the three gluon
vertex
 \begin{eqnarray}\label{3geff}
 \tilde{g}(k_1^2,k_2^2,k_3^2)&\equiv& g_0(1+\tilde{A}_0(k_1^2,k_2^2,k_3^2))\nonumber\\
 \tilde{\a}(a,b,c) &\equiv& {\tilde{g}^2(a,b,c)\over 4\pi}\approx {\a_0\over
 1-2\tilde{A}_0(a,b,c)},
 \end{eqnarray}
 first obtained by Lu in \cite{Lu3gg}.
 Our amplitude then takes the final form
 \bege\label{Mampfinal}
 {\cal M} = {\cal C} g(k_1^2)g(k_2^2)g(k_3^2)\tilde{g}(k_1^2,k_2^2,k_3^2)
 {V_1^{\mu_1}V_2^{\mu_2}V_3^{\mu_3}\over k_1^2 k_2^2 k_3^2}
  \Big[ \hat{a}_0 + A_+ \hat{a}_+ + A_- \hat{a}_- + H\hat{h}\Big].
 \ende
 Recall from the previous section that $A_{\pm},H \propto N_c-N_f$
 in QCD.

The three-gluon effective coupling evolves according to
 \bege
 \tilde{\a}(a,b,c) = {\tilde{\a}(a_0,b_0,c_0)\over 1-2\Big(\tilde{A}_0(a,b,c)-\tilde{A}_0(a_0,b_0,c_0)\Big)}
 \ende
In four dimensions with regularization scheme $R=DRED \;{\rm or}\;
DREG$ we have
 \begin{eqnarray}\label{Atilde}
 \tilde{A}_0(a,b,c) &=& -{\a_s\over 8\pi}\b_0\Big[ L(a,b,c)-\log{\mu^2}-C_{UV} -\eta_3 \Big]
 \nonumber\\
 {\rm where} \;\; \b_0 &=& {11\over 3}N_c-{2\over 3}N_f-{1\over 6}N_s
 \nonumber\\
 C_{UV} &=& {1\over \e} -\g_E +\log{4\pi}
 \\
 \eta_3 &=& (2+\Omega) + {N_c\over 3\b_0}\d_{R,DREG}
 \nonumber\\
 \Omega &=& {16\over 3\sqrt{3}}{\rm Cl}_2(\pi/3)\approx 3.125
 \nonumber
 \end{eqnarray}
 The scheme dependence $\d_{R,DREG}$ is explained in more detail in
 Appendix D.
 Here we have defined
 \bege\label{Labc}
 L(a,b,c) = {1\over \CK}\Big( \a\g\log{a}+\a\b\log{b}+\b\g\log{c}-abc \bar{J}(a,b,c) \Big) +
 \Omega,
 \ende
 and the (massless) triangle integral function $\bar{J}=\bar{J}(a,b,c)= -16i\pi^2 J(a,b,c)$
 is given in Appendix B in terms of Clausen functions,
 Eqs.(\ref{JKp},\ref{JKm}).
 This result (Eqs.(\ref{3geff},\ref{Atilde})) differs from Lu \cite{Lu3gg} by only the finite constants
 which (slightly) affects the numerical extraction from data. The discrepancy
 can be traced to the inconsistent application of dimensional
 regularization in \cite{Lu3gg}.

The logarithm-like function $L$ satisfies
 \bege
 L(a,a,a)=\log{a}
 \ende
since
 \bege
 \bar{J}(a,a,a)={4\over a\sqrt{3}} \Cl2(\pi/3).
 \ende
 One can use the real part of this function to define an effective
 scale of the three-gluon vertex:
 \begin{eqnarray}
 L(a,b,c) &=& \log{\left( Q_{eff}^2(a,b,c)\right) } + i \Im L(a,b,c) \nonumber\\
 Q_{eff}^2(a,b,c) &=&
 |a|^{\a\g/\CK}|b|^{\a\b/\CK}|c|^{\b\g/\CK}
 \exp{\Big( \Omega-{abc\over \CK}\Re\bar{J}(a,b,c)\Big)}.
\end{eqnarray}
This is sensible since the dimensions of $Q_{eff}^2(a,b,c)$ are
indeed mass squared.

\begin{figure}[htb]
 \centering \includegraphics[height=2.5in]{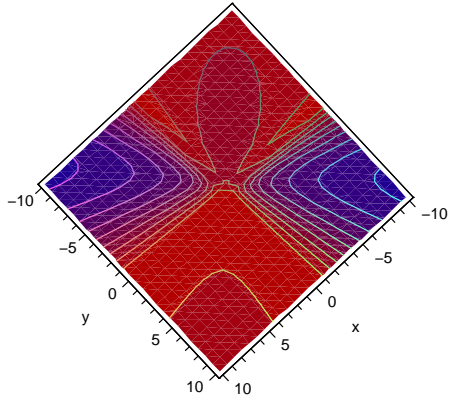}
 \caption[*]{ A contour plot of $Q^2_{eff}(1,x,y)$. The contours, from red to blue, are at
 $2,4,6,8,10,12,14,16,18,20$. }
 \label{fig:5}
\end{figure}
\begin{figure}[htb]
 \centering \includegraphics[height=2.5in]{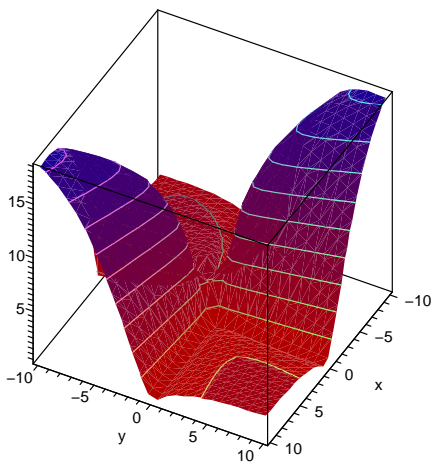}
 \caption[*]{A three-dimensional contour plot of $Q^2_{eff}(1,x,y)$. }
 \label{fig:6}
\end{figure}

The three gluon effective charge $\tilde{\a}(a,b,c)$ is related to
the usual $\bar{MS}$ coupling $\bar{\a}(q^2)$ by
 \begin{eqnarray}\label{3gbar}
 {1\over \tilde{\a}(a,b,c)} &=& {1\over\bar{\a}(\mu^2)} +{\b_0\over 4\pi}
 \Big( \log{Q_{eff}^2(a,b,c)\over\mu^2} + i\Im{L(a,b,c)} -\eta_3 \Big) \nonumber\\
 &=& {1\over\bar{\a}(e^{-\eta_3}Q_{eff}^2(a,b,c))} +i{\b_0\over 4\pi}\Im{L(a,b,c)}.
 \end{eqnarray}
 Since $\exp{\left( \eta_3\over 2\right) }\approx 14$, we see that when using $\bar{MS}$,
 the scale should be {\it fourteen} times lower than the typical
 virtualities of the gluons, given by $Q_{eff}(a,b,c)$. Of
 course, this is true only if the three-gluon vertex diagram dominates the
 physical process. In general there will be different scales at the
 various quark-gluon vertices when using the PT scheme (as seen in
 Eq.(\ref{Mampfinal})).
 In contrast, in $\bar{MS}$ the same scale is used at every vertex.
 The following approximate values of the three-gluon coupling are derived from
 Eq.(\ref{3gbar}) (including the effects of quark masses which discussed in the next section)
 for various symmetric timelike(T) and spacelike(S)
 configurations :
$$
 \begin{array}{ll}
  {\rm SSS}: \;\;\;\;\;\;\; & \tilde{\a}(-M_Z^2,-M_Z^2,-M_Z^2) \approx 0.192
 \\
  {\rm SST}: \;\;\;\;\;\;\; & \tilde{\a}(-M_Z^2,-M_Z^2,+M_Z^2) \approx 0.157+0.023I
 \\
  {\rm STT}: \;\;\;\;\;\;\; & \tilde{\a}(-M_Z^2,+M_Z^2,+M_Z^2) \approx 0.156+0.025I
 \\
  {\rm TTT}: \;\;\;\;\;\;\; & \tilde{\a}(+M_Z^2,+M_Z^2,+M_Z^2) \approx 0.170+0.062I
 \end{array}
$$
It is clear that the three-gluon coupling is stronger than naively
expected from $\a_{\bar{MS}}(M_Z) \approx 0.118$.

The effective scale $Q_{eff}^2(a,b,c)$ satisfies the following
relations:
\begin{eqnarray}
 Q_{eff}^2(a,b,c) &=& Q_{eff}^2(-a,-b,-c)\nonumber\\
 Q_{eff}^2(\l a,\l b,\l c) &=& |\l| Q_{eff}^2(a,b,c)\nonumber\\
 Q_{eff}^2(a,a,a) &=& |a| \\
 Q_{eff}^2(a,-a,-a) &\approx& 5.54|a| \nonumber\\
 Q_{eff}^2(a,a,c) &{\approx}& |c| e^{\Omega - 2} \;\;\;\;\;\;\; {\rm for}\;\; |a| \gg |c| \nonumber\\
 Q_{eff}^2(a,-a,c) &{\approx}& |c| e^{\Omega} \;\;\;\;\;\;\; {\rm for}\;\; |a| \gg |c| \nonumber\\
 Q_{eff}^2(a,b,c) &{\approx}& {|b||c| \over |a|} e^{\Omega} \;\;\;\;\;\;\; {\rm for}\;\; |a|\gg |b|,|c|.
 \nonumber
\end{eqnarray}
 Lu \cite{Lu3gg} has previously found the last of these limits in the case
 where all momenta are spacelike, giving an effective scale
 $Q_{min}Q_{med}/Q_{max}$.
It should be noted that the rate of convergence to the above limits
strongly depends on the signatures (${\rm S}\equiv{\rm
spacelike}\leftrightarrow p^2<0$, ${\rm T}\equiv{\rm
timelike}\leftrightarrow p^2>0$) of the virtualities $a,b,c$. If the
signatures are mixed (TTS) or (TSS) then the convergence is very
slow, and the effective scale tends to stay larger compared to the
cases (SSS) or (TTT).

Some plots demonstrating the novel behavior of $Q_{eff}^2$ are given
in Figs.(\ref{fig:3},\ref{fig:4},\ref{fig:5},\ref{fig:6}).

\section{Phenomenological Effects of Internal Masses}

So far, all fields propagating in the triangle graphs have been
treated as massless. This was useful for simplifying the
discussion and elucidating the general structure of the radiative
corrections and the $N=4$ sum rules. However, in real world
applications one usually does not have all three gluon
virtualities in the same desert region
 $M_i \ll a,b,c \ll M_{i+1}$. Thus, mass corrections should be
taken into account. We have calculated the effects of massive
fermions (MQ), massive scalars (MS), and massive gauge bosons (MG)
for all of the form factors; the complete results are given in
Appendix E. The corrections for the case of massive fermions were
first obtained in Ref.\cite{Davydychev:2001uj} and we are in
agreement.
 Here we will focus on the massive quark ($MQ$) contribution
 to the form factor multiplying the tree level tensor
 structure, which from Appendix E and section 3 is
 \bege\label{A0MQ}
 A_0(MQ) ={4M^2\over 3(d-2)}J_M +{3d-8\over 6\CK(d-2)}\Big[ abcJ_M + a\b J_{1M} + b\g J_{2M} + c\a J_{3M}
 \Big].
 \ende
 Here $J_M$, $J_{1M}$, $J_{2M}$, and $J_{3M}$
 are the massive analogs of $J,J_1,J_2$ and $J_3$, respectively.
 The two-point function $J_{1M}$ and tadpole $T_M$ are reviewed in Appendix A, while
 Appendix B is devoted to a discussion of the massive triangle integral, $J_M$, and its analytic
 continuations and various limits.

\begin{figure}[htb]
 \centering \includegraphics[height=2.5in]{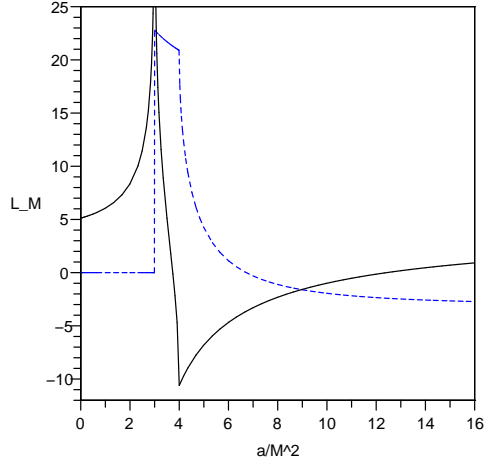}
 \caption[*]{$L_{MQ}(a/M^2,a/M^2,a/M^2)$ vs. $a/M^2$ for timelike $a>0$. The solid line is the real part
 and the dashed line is the imaginary part.     }
 \label{fig:7}
\end{figure}
\begin{figure}[htb]
 \centering \includegraphics[height=2.5in]{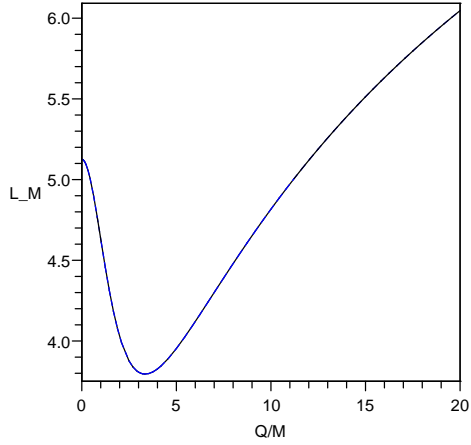}
 \caption[*]{$L_{MQ}(-Q^2/M^2,-Q^2/M^2,-Q^2/M^2)$ vs. $Q/M$ for spacelike $-Q^2<0$. }
 \label{fig:8}
\end{figure}

 As in the previous section, when considering a physical matrix element we always have
 the combination $\tilde{A}_0 = A_0 -\half\big(\Pi_1+\Pi_2+\Pi_3\big)$ multiplying the tree-level tensor
 structure. This leads us to consider the massive quark
 contribution
 $\tilde{A}_0(MQ) = A_0(MQ)
 -\half\big(\Pi_1(MQ)+\Pi_2(MQ)+\Pi_3(MQ)\big)$, which upon using Eq.(\ref{PiM}),
 inserting the prefactor $ig^2$, and expanding around $d=4$
 becomes
 \begin{eqnarray}
 \tilde{A}_0(MQ) &=& -{\a_s\over 4\pi}\Bigg[ {1\over 3} \left(C_{UV}-\log{M^2\over \mu^2}\right)+{2\over 3}
 +{1\over 3\CK} \left( abc \bar{J_M}-\a\g \CL(a)-\a\b\CL(b)-\b\g\CL(c) \right) \nonumber\\
 &+& {2M^2\over 3}\left(\bar{J_M} +{2-\CL(a)\over a}+{2-\CL(b)\over b}+{2-\CL(c)\over c} \right)
 \Bigg].
 \end{eqnarray}
Here $\bar{J_M}=-16i\pi^2 J_M$ and
 \bege
 \CL(a) = v(a)\log{v(a)+1\over v(a)-1} \;\;{\rm where}\;\; v(a)=\sqrt{1-{4(M^2-i\e)\over a}},
 \ende
 comes from $J_{1M}$ and has the analytic continuations given in Eq.(\ref{J1M}).
 The three-scale logarithm-like function for massive quarks (MQ) is thus given by
 \begin{eqnarray}
 L_{MQ} \left({a\over M^2},{b\over M^2},{c\over M^2} \right) &=&
 {1\over \CK}\Big( \a\g\CL(a)+\a\b\CL(b)+\b\g\CL(c)-abc\bar{J_M}(a,b,c) \Big) + \Omega
 \nonumber\\
 &+& 2M^2\left( {\CL(a)-2\over a} + {\CL(b)-2\over b} + {\CL(c)-2\over c}-\bar{J_M} \right)
 \\
 \tilde{A}_0(MQ) &=& \left({\a_s\over 12\pi}\right)
 \left[  L_M \left({a\over M^2},{b\over M^2},{c\over M^2} \right)
 + \log{M^2\over \mu^2} - C_{UV} -(2+\Omega)\right] \nonumber.
 \end{eqnarray}
This massive logarithm-like function has the following limits :
 \begin{eqnarray}
 L_{MQ} \left({a\over M^2},{b\over M^2},{c\over M^2} \right)&\approx& 2+\Omega
 \;\;\;\;\;\;\;\;\;\;\;\;\;\;\;\;\;\;\;\;\;\; M^2\gg |a|, |b|, |c|\nonumber\\
 L_{MQ} \left({a\over M^2},{b\over M^2},{c\over M^2} \right)&\approx&
 L(a,b,c) -\log{M^2}
 \;\;\;\;\;\;\; M^2\ll |a|, |b|, |c|,
 \end{eqnarray}
 with the number $\Omega\approx 3.125$ defined in Eq.(\ref{Atilde}).
The convergence to the massless limit is very slow, indicating
that threshold effects must be included for most applications.

\begin{figure}[htb]
 \centering \includegraphics[height=2.5in]{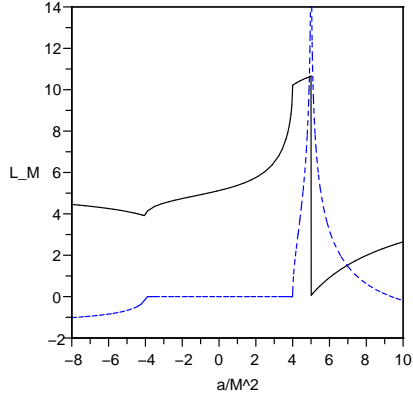}
 \caption[*]{$L_{MQ}(a/M^2,a/M^2,-a/M^2)$ vs. $a/M^2$. The solid line is the real part
 and the dashed line is the imaginary part. The real thresholds are at $a=\pm 4M^2$
 while the pseudo-threshold is at $a=5M^2$.}
 \label{fig:9}
\end{figure}

In Figs.(\ref{fig:7},\ref{fig:8}) we have plotted $L_{MQ}$ for the
symmetric case $a=b=c$ for timelike and spacelike momenta. For the
timelike case, the threshold at $a=4M^2$ and the pseudo-threshold
at $a=3M^2$ are evident. In Fig.(\ref{fig:9}) the mixed case (TTS)
$L_{MQ}(a/M^2,a/M^2,-a/M^2)$ is plotted, where the thresholds at
$a=\pm 4M^2$ and the pseudo-threshold at $a=5M^2$ are evident. For
the purely timelike (TTT) case in Fig.(\ref{fig:7}),  there is a
discontinuity in the imaginary part and the real part diverges at
the pseudo-threshold. In contrast, for the mixed signature case
(TTS) the imaginary part diverges while the real part is
discontinuous. This pseudo-threshold phenomena is explained in
more detail in Appendix B.

\begin{figure}[htb]
 \centering \includegraphics[height=2.5in]{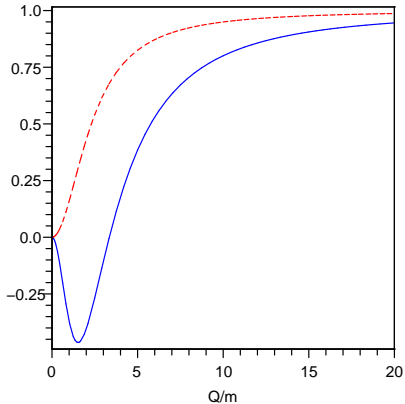}
 \caption[*]{{\bf The effective number of quark flavors.} The lower blue solid curve is
 $N_F(Q^2/M^2)$ for the symmetric spacelike ($a=b=c=-Q^2$) three-gluon vertex, while the
 upper dashed red curve is the fermion number of flavors
 $N_{1/2}(Q^2/M^2)$ for the single scale effective charge.  }
 \label{fig:10}
\end{figure}

From the above results, one can define the effective number of
active quarks which characterizes the effects of quark mass :
 \bege\label{NF3}
 N_F \left({a\over M^2},{b\over M^2},{c\over M^2} \right)
 =-M^2{d\over dM^2} L_{MQ} \left({a\over M^2},{b\over M^2},{c\over M^2}
 \right).
 \ende
This clearly goes to zero and one in the limits
 $M^2\gg |a|, |b|, |c|$ and  $M^2\ll |a|, |b|,|c|$, respectively.

 To motivate this definition, let's look at the
single-scale pinch-technique effective charge (using the notation
of \cite{Binger:2003by}) as a function of spacelike momenta
$a=-Q^2$
 \bege
 {1\over \tilde{\a}(Q^2)} = {1\over \a_0} +{1\over 4\pi}\sum_p
 \b_p\left( L_p(Q^2/m_p^2) - \log{\mu^2\over m_p^2} -C_{UV} -\eta_p
 \right),
 \ende
where $\b_p$ is the contribution of each particle $p$ to the first
$\b$ function coefficient, and to a very good numerical
approximation (for spacelike momenta)
 \bege
 L_p(Q^2/m_p^2) \approx \log{\left( e^{\eta_p} + {Q^2\over m_p^2} \right)},
 \ende
where the constants $\eta_p$ are $5/3,8/3,$ and $40/21$ for
massive fermions, scalars, and gauge bosons, respectively. The
exact one-loop formula are given in Eq.(23-26) of
Ref.\cite{Binger:2003by}, although the analytic continuation in
Eq.(26) of that paper should have opposite imaginary part. This
effective charge satisfies the RGE
 \begin{eqnarray}\label{Np}
 {d\tilde{\a}(Q^2)\over d\log{Q^2}} &=& -{\tilde{\a}^2\over 4\pi}
 \sum_p \b_p {d L_p(Q^2/m_p^2)\over d\log{Q^2}}
 \equiv-{\tilde{\a}^2\over 4\pi}
 \sum_p \b_p N_p\left( {Q^2\over m_p^2} \right)
 \nonumber\\
 N_p\left( {Q^2\over m_p^2} \right) &\equiv& {d L_p(Q^2/m_p^2)\over
 d\log{Q^2}} =-{d L_p(Q^2/m_p^2)\over d\log{m_p^2}} \approx {1\over 1+{m_p^2\over Q^2}e^{\eta_p}}.
 \end{eqnarray}
The function $N_p$ goes to one when $Q^2\gg m_p^2$ and zero when
$Q^2\ll m_p^2$ and unambiguously measures what fraction of particle
$p$ is ``turned on" at scale $Q^2$.

Moving back to the three-scale case, we now have the complication
that our effective charge is a solution of a multi-scale RGE
 \bege
 {d\tilde{\a}(a,b,c)\over d\log{a}} = -{\tilde{\a}^2\over 4\pi}
 \sum_p \b_p {d \over d\log{a}} L_{MQ}\left({a\over M^2},{b\over M^2},{c\over
 M^2}\right),
 \ende
and two other permutations with $a\raw b$ or $a\raw c$. This leads
to three different $N_f$'s :
 \bege
 N_f \left({a\over M^2}\Big|{b\over M^2},{c\over M^2} \right)
 = {d \over d\log{a}} L_{MQ}\left({a\over M^2},{b\over M^2},{c\over
 M^2}\right),
 \ende
and two cyclic permutations, each of which goes to $1/3$ in the
symmetric desert $a=b=c\gg M^2$. This suggests adding all three
together to define a symmetric
 \bege
 N_F \left({a\over M^2},{b\over M^2},{c\over M^2} \right) =
 \left( {d \over d\log{a}}+{d \over d\log{b}}+{d \over d\log{c}}\right)
 L_{MQ}\left({a\over M^2},{b\over M^2},{c\over M^2} \right),
 \ende
 which is in fact the same as given in Eq.(\ref{NF3}).

The results for $N_F$ can be obtained with the help of the results
in Appendix B. Instead of presenting these lengthy results, let us
focus on the symmetric case $a=b=c$, where we find for spacelike
$a<0$
 \begin{eqnarray}
 N_F\left( {a\over M^2} \right)
 &\equiv& {d \over d\log{a}} L_{MQ}\left({a\over M^2},{a\over M^2},{a\over M^2}\right)
    = -{d \over d\log{M^2}} L_{MQ}\left({a\over M^2},{a\over M^2},{a\over M^2}\right)
 \nonumber\\
 &=& 1+18{M^2\over a} +2M^2\bar{J_M}(a,a,a) + 54M^2\CL(a) {a-2M^2\over (a-3M^2)(a-4M^2)}.
 \end{eqnarray}
In this example, spacelike $a$ is chosen to avoid the
pseudo-threshold $a=3M^2$ and the threshold $a=4M^2$.
Fig.(\ref{fig:10}) shows a plot of this along with single-scale
quark number of flavors function $N_{s=1/2}$ from Eq.(\ref{Np}).

The negative value of $N_F$ at $0\lsim Q \lsim 4M$ is not entirely
novel, as a similar behavior was also found in the context of
two-loop quark mass corrections to V-scheme effective charge
\cite{Brodsky:1999fr}. It is essentially due to the anti-screening
of color charge, this case in the triangle interaction, and does not
arise in the one-loop single scale effective charge, as seen in
Fig.(\ref{fig:9}).

Using the results of Appendix E, the above analysis can be easily
extended to the case of massive scalars (MS) or massive gauge bosons
(MG), which have the following logarithm-like functions
 \begin{eqnarray}
 L_{MS} \left({a\over M^2},{b\over M^2},{c\over M^2} \right) &=&
 {1\over \CK}\Big( \a\g\CL(a)+\a\b\CL(b)+\b\g\CL(c)-abc\bar{J_M}(a,b,c) \Big) + \Omega
 \nonumber\\
 &-& 4M^2\left( {\CL(a)-2\over a} + {\CL(b)-2\over b} + {\CL(c)-2\over c}-\bar{J_M} \right)
 \\
 L_{MG} \left({a\over M^2},{b\over M^2},{c\over M^2} \right) &=&
 {1\over \CK}\Big( \a\g\CL(a)+\a\b\CL(b)+\b\g\CL(c)-abc\bar{J_M}(a,b,c) \Big) + \Omega
 \nonumber\\
 &+& {4\over 7}M^2\left( {\CL(a)-2\over a} + {\CL(b)-2\over b} + {\CL(c)-2\over c}-\bar{J_M}
 \right).
 \end{eqnarray}
 The qualitative behavior is the same as the quark case.

Finally, we should consider the limitations of the effective scale
$Q^2_{eff}(a,b,c)$ introduced in the last section and effective
number of flavors $N_F(a/M^2,b/M^2,c/M^2)$ discussed in this
section. Given the complicated structure of the full mass dependent
form factors, such tools for characterizing and understanding the
behavior of the vertex are helpful. However, in a real calculation
such methods may be of limited use and the full mass dependent
results should be used. For example, the effective scale $Q^2_{eff}$
has been defined only in the massless case so far because the
definition becomes complicated and somewhat arbitrary in the massive
case. In particular, consider the possible definition (for QCD):
 \begin{eqnarray}
 &\Re& \!\!\! \Bigg[ {11\over 3}C_A L(a,b,c) - {2\over 3}\sum_q \Bigg(
 L_{MQ}\left({a\over M_q^2},{b\over M_q^2},{c\over M_q^2}\right)
 +\log{M_q^2}\Bigg)\Bigg] \nonumber\\
 &\equiv&  \Bigg( {11C_A\over 3} - {2\over 3}\sum_q
 \tilde{N_q}\left({a\over M_q^2},{b\over M_q^2},{c\over M_q^2}\right) \Bigg)
 \log{\tilde{Q}^2_{eff}(a,b,c)},
 \end{eqnarray}
where $\tilde{N_q}$ is some suitably defined number of flavors,
possibly a step function such as $\d(a+b+c-3M_q^2)$, possibly the
$N_F$ defined in Eq.(\ref{NF3}), or some other definition. It should
be clear that any choice of $\tilde{N_q}$ determines
$\tilde{Q}_{eff}^2$, and vice versa, and there seems to be no
compelling choice for these quantities. Furthermore, in the approach
advocated here, the couplings at each vertex depend on physical
momentum scales which will typically be integrated over in the phase
space. Thus, matching onto a conventional $\bar{MS}$ type approach
can only be done at the end of the calculation, so that trying to
define a $Q_{eff}^2$ at an intermediate stage is not very useful.

Thus, in real world applications, one should generally use the full
results for the mass dependent form factors. This constitutes a
multi-scale analytic renormalization scheme that contains
information which cannot be obtained in the simple single-scale
leading-log renormalization methods. In other words, every
three-gluon vertex (at tree-level) can be dressed, or ``RG
improved", with this gauge-invariant effective coupling and the
associated form factors, which are process independent and contain
more information than the $\bar{MS}$ procedure.

\section{Conclusions and Future Directions}

 The results of this paper represent only a fraction of what is
 needed for a re-organization of perturbation theory into fully gauge-invariant
 pieces with physical content, each of which can be renormalized independently,
 leading naturally to a physical multi-scale analytic renormalization scheme.
 This is possible due to the remarkable properties of the pinch technique (PT)/
 Background Field Method quantum Feynman gauge (BFMFG) Green's
 functions. There is still much progress that can be made in
 calculating these Green's functions in perturbation theory.

 The present paper gives a complete and general characterization
 of the off-shell three-gluon vertex at one-loop. A similar study of
 the gauge-invariant triple gauge boson vertices of the Standard Model
 \cite{Papavassiliou:1995fw} would be very
 useful. It may also be possible to quantitatively look at
 the unification of triple gauge boson vertices and couplings, in
 analogy with the work on the unification of single-scale PT
 couplings\cite{Binger:2003by}. Some progress has been made on the
 conventional gauge-dependent three-gluon vertex at two loops
 \cite{Davydychev:1998aw}\cite{Davydychev:1997vh},
 which gives hope for eventually treating the gauge-invariant
 three-gluon vertex at two-loops.

 The gauge boson two-point functions and the associated effective
 charges for QCD \cite{Cornwall:1981zr}\cite{Watson:1996fg}, electroweak theory
 \cite{Papavassiliou:1989zd}\cite{Degrassi:1992ue}\cite{Papavassiliou:1996fn}, and supersymmetric grand unified
 models \cite{Binger:2003by} have been calculated in the past to one-loop.
 We know from general principles the divergent parts at two-loops,
 but no complete two-loop PT calculation as yet exists.
 To fill this gap, the two-loop gluon PT self-energy will be presented
 in the near future \cite{2loopPTSE}. This will allow for a more
 precise determination of coupling from data, as well as giving the
 three-loop $\b$ function coefficient. Furthermore, by the Ward
 identity in Eq.(\ref{wardid}), this also yields the longitudinal
 form factors of $\Gamma_{\mu_1\mu_2\mu_3}^{(ggg)}$
 through two loops.

 The gauge-invariant PT/BFMFG quark self-energy turns out to be
 equal to the conventional self-energy in the Feynman gauge \cite{Binosi:2001hy},
 and so is known through two-loops \cite{Fleischer:1998dw}. Due to
 the Ward identity \cite{Papavassiliou:1999bb} satisfied by the PT/BFMFG quark-gluon
 vertex, this also yields the longitudinal form factors of $\Gamma_{\mu}^{(qqg)}$
 through two loops.

 In QCD, another logical step is the four-gluon vertex at
 one-loop. In the general off-shell case, there are hundreds of independent
 tensors structures and form factors.

Beyond perturbation theory, the study of Dyson-Schwinger Equations
\cite{Alkofer:2000wg} and renormalons \cite{Beneke:1998ui} in the
PT/BFMFG approach may yield new insight.

To summarize, in this paper we have analyzed the behavior of the
gauge-invariant three-gluon vertex at one-loop. Starting from the
symmetry principles governing the vertex, a convenient tensor basis
decomposition was given in Eqs.(\ref{pmbasis},\ref{sympmdef}). As
seen in Eq.(\ref{GammaOS}) and the subsequent discussion, this basis
is the most convenient for phenomenological studies, since it is
built out of ``transverse" ($-$) and ``longitudinal" ($+$) momenta,
the latter of which vanish when dotted into external on-shell
vertices, thus leading to relatively simple matrix elements. In the
case considered in section 4, only four form factors remain, rather
than the thirteen which would be present in a generic basis.
Nonetheless, the choice of basis is only a matter of convenience,
and the real physics lies in the thirteen non-vanishing form factors
given explicitly in section 3.

The supersymmetric relations between the scalar, quark, and gluon
contributions leads to a simple presentation of the results for a
generic (unbroken) gauge theory. Only the gluon contributions to the
form factors are given explicitly in section 3, while the quark and
scalar contributions are inferred from the homogeneous relation
$F_G+4F_Q+(10-d)F_S=0$ and the results for the relatively simple
sums $\Sigma_{QG}(F) \equiv {(d-2)\over 2}F_Q + F_G$ which are given
in section 3 for each form factor $F$. The extension to the case of
internal masses is outlined in Appendix E and leads to the modified
sum rule $F_{MG}+4F_{MQ}+(9-d)F_{MS}=0$.

The phenomenology is largely determined by the form factor of the
tree-level tensor structure, which in section 4 is used to define a
three-scale effective charge $\tilde{\a}(a,b,c)$. In addition, the
characteristic scale $Q^2_{eff}(a,b,c)$ governing the behavior of
the vertex and the effective charge was analyzed, thus providing a
natural extension of BLM scale setting \cite{Brodsky:1982gc} to the
three-gluon vertex. Physical momentum scales always set the scale of
the coupling. The phenomenological effects of quark masses are
discussed in section 5 and are found to be important for generic
physical applications, since decoupling is slow and a complicated
threshold and pseudo-threshold behavior is observed. An important
next step is to fully apply these techniques to a physical process.
In the future we will present such an analysis for the hadronic
production of heavy quarks, where the importance of the form factors
other than the tree-level one ($A_0$) will be addressed. The
interpretation of the pseudo-threshold phenomena also deserves
further study.

{\center{\Large \bf Acknowledgements}}
M.B. would like to thank Lance Dixon for useful discussions
regarding the second order formalism of the BFM and the
supersymmetric decomposition of one-loop amplitudes.

\newpage
{\center{\Large \bf Appendix A: Reduction to Scalar Integrals}}
\vspace{0.75cm}

First, we will describe the evaluation of the
massless integrals, and then briefly mention the modifications due
to internal masses. As before, we will use the shorthand notation
 \begin{eqnarray}
 a = p_1^2 \;\;\;\;\; b=p_2^2 \;\;\;\;\; c=p_3^2 \;\;\;\;\;
 \a = p_1 \!\cdot\! p_2 \;\;\;\;\; \b = p_2 \!\cdot\! p_3 \;\;\;\;\; \g = p_3 \!\cdot\! p_1
 \end{eqnarray}

In order to evaluate the integrals in an efficient manner, it is
very convenient to choose a manifestly symmetric routing of the
loop momenta, as shown in Fig.\ref{fig:1}, where clearly
 \bege\label{loopmom}
 l_1=p_2+l_3 \;\;\;\; l_2=p_3+l_1 \;\;\;\; l_3=p_1+l_2.
 \ende
Of course there is only one integration momenta $l$, which can be
chosen to be $l_1, l_2,$ or $l_3$, thus breaking the cyclic
symmetry. However, using the symmetric labeling greatly simplifies
the analysis.

First we decompose the full vertex $\Gamma$ into longitudinal ($L$)
and transverse ($T$) parts, $\Gamma=\Gamma_L+\Gamma_T$, as in
Eq.(\ref{LTbasis}). The tensor integrals in Eq.(\ref{GQSdef}) are
then converted into scalar integrals by applying projection
operators. In doing so, the longitudinal ($L$) and transverse ($T$)
parts essentially decouple, and the ten independent $L$ form factors
are easily found either directly, or by solving the Ward ID,
resulting in Eq.(\ref{LFFs}). The remaining four $T$ parts are found
by applying the following four projection operators to
Eq.(\ref{GQSdef}) : $200,030,001,$ and $231$, where as in Table 1 we
have defined $030\equiv p_{3\mu_2}g_{\mu_1\mu_3}$, etc. Thus, for
the gluon contribution $G$ we have four scalar integrals :
$G(200)\equiv p_{2\mu_1}g_{\mu_2\mu_3}G_{\mu_1\mu_2\mu_3}, G(030),
G(001),$ and $G(231)$. Similarly, there are four integrals for the
quarks and scalars as well. In the numerator of each of these
integrals there will be various dot products of momenta, which can
always be reduced to momentum squares using Eq.(\ref{loopmom}). For
example,
 $p_1 \!\cdot\! l_2 = (l_3^2-l_2^2-p_1^2)/2$ and $l_1 \!\cdot\! l_3 =
(l_1^2+l_3^2-p_2^2)/2$. Thus we are left with integrals of the form
 \bege
  I_{ijk} = \int {(l_1^2)^i (l_2^2)^j (l_3^2)^k \over
  l_1^2l_2^2l_3^2},
 \ende
 where $\int \equiv \int {d^dl\over (2\pi)^d}$ and $i,j,k\in \{0,1,2\}$.
 Using the standard rules of dimensional regularization, it is easy
 to see that any integral with any two of $i,j,k$ nonzero must
 vanish. Furthermore, it is straightforward to show that
 \bege
 I_{200}=-\b I_{100} \;\;\;\;\; I_{020}=-\g I_{010} \;\;\;\;\; I_{002}=-\a
 I_{001}.
 \ende
 Thus we are left with only two types of integrals:
 (1) the trivial two point integrals $J_1$, $J_2$, and $J_3$, where
 \bege
 J_1= I_{100}=\int {1\over l_2^2 l_3^2} = \int {1\over l^2
 (l+p_1)^2},
 \ende
 and (2) the master triangle integral
 \bege
 J \equiv J(p_1^2,p_2^2,p_3^2) = \int {1\over l_1^2 l_2^2 l_3^2}.
 \ende

For the gluon contribution, for example, one then has a system of
four equations with four unknowns, the transverse form factors.
Denoting the gluon contribution to the longitudinal projections by
$L_G(200)=200\cdot \Gamma_L(G)$, etc. we solve for the transverse
form factors
 \begin{eqnarray}
 \left( \begin{array}{cccc}
 \bar{F}_{12}(G) \\
 \bar{F}_{23}(G) \\
 \bar{F}_{31}(G) \\
 \bar{H}(G)
 \end{array} \right) &=& M_T^{-1}
 \left( \begin{array}{cccc}
 G(200)-L_G(200)\\
 G(030)-L_G(030) \\
 G(001)-L_G(001) \\
 G(231)-L_G(231)
 \end{array} \right)
 \nonumber\\
 {\rm where} \;\;\;\;\; M_T &=& -\CK
 \left( \begin{array}{cccc}
  \b & (d-1)\b & \b & 2-d \\
  \g &\g & (d-1) \g & 2-d \\
  (d-1)\a & \a & \a & 2-d \\
  \CK & \CK & \CK & 0
 \end{array} \right),
\end{eqnarray}
and similarly for the quark and scalar contributions.

The above procedure can also be followed for the massive case, with
only a few modifications. First, the tadpole $M^2T_M = \int {1\over
l_i^2 - M^2}$ does not vanish.  Thus, instead of
 $\int {l_1^2\over l_2^2 l_3^2} = -\b J_{1}$ we now have
 \bege
 \int {l_1^2-M^2\over (l_2^2-M^2)(l_3^2-M^2)} = -\b J_{1M} + M^2T_M,
 \ende
where $J_{1M} = \int{1\over (l_2^2-M^2)(l_3^2-M^2)} = \int{1\over
(l^2-M^2)((l+p_1)^2-M^2)}$. We also need the following result and
permutations :
 \bege
 \int{l_2^2-M^2\over l_3^2-M^2} = \int{l_3^2-M^2\over
 l_2^2-M^2} = aM^2T_M.
 \ende
Finally, we have the master triangle integral with nonzero masses
 \bege
 J_M \equiv J_M(p_1^2,p_2^2,p_3^2) = \int {1\over (l_1^2-M^2) (l_2^2-M^2) (l_3^2-M^2)}.
 \ende

To summarize, in the massive case we need $J_M$, $J_{1M}$, $J_{2M}$,
$J_{3M}$, and $T_M$. In the massless case we need $J$, $J_1$, $J_2$,
and $J_3$. For each of these we pull out the factor ${i\over
16\pi^2}$ and define $J_M = {i\over 16\pi^2} \bar{J_M}$, etc.

Some formula for these integrals in $d$ dimensions can be found in
\cite{Davydychev:1996pb}. Here we will give only the expansions in
four dimensions and define $C_{UV} = {1\over \e} -\g_E +
\log{4\pi}$ where $d=4-2\e$.

The tadpole integral is
 \bege
 \bar{T_M} = C_{UV} + 1  - \log{M^2\over \mu^2}.
 \ende

The two point integral is
 \begin{eqnarray}\label{J1M}
 \bar{J_{1M}} &=& C_{UV} + 2  - \log{M^2\over \mu^2} - \CL(a) \\
 \CL(a) &=&
 \left\{ \begin{array}{lll}
  2v \tanh^{-1}{(v^{-1})} = v\log{v+1\over v-1} \\
  2\bar{v} \tan^{-1}{(\bar{v}^{-1})} \\
  2v \tanh^{-1}{(v)} -i\pi v = v\log{1+v\over 1-v} -i\pi v
 \end{array} \right\}
 \;\;{\rm for} \;\; \left\{ \begin{array}{lll}
 a<0 \\ 0<a<4M^2 \\ a>4M^2 \end{array} \right\}
 \nonumber
 \end{eqnarray}
and the generalized velocities are
 \bege
 v = \sqrt{1-{4M^2\over a}} \;\;\;\;\;\;\;\;\;\;\;\; \bar{v} = \sqrt{{4M^2\over a}-1}.
 \ende

 In the massless limit this becomes
 \bege
 \bar{J_1} = C_{UV} + 2 - \log{|a|\over \mu^2}+i\pi \theta(a).
 \ende

\vspace{0.75cm}
{\center{\Large \bf Appendix B: Results for the Triangle Integral}}
\vspace{0.75cm}

 The massive triangle integral
 \bege
 J_M \equiv J_M(p_1^2,p_2^2,p_3^2) = \int{d^4l\over (2\pi)^4} {1\over (l_1^2-M^2+i\e) (l_2^2-M^2+i\e) (l_3^2-M^2+i\e)}
 \ende
 is finite in four dimensions. We will give the results for $\bar{J_M} = -i 16\pi^2
 J_M$. This integral has been discussed previously in the literature
 \cite{'tHooft:1978xw}\cite{vanOldenborgh:1989wn}\cite{Davydychev:1997wa}\cite{Davydychev:2001uj}.
 In particular, 'tHooft and Veltman \cite{'tHooft:1978xw} derived a
 formula which is valid for all values of
 the kinematic variables $a,b,c$ and mass $M$, although careful
 analytic continuation is required. We will first write the results
 of \cite{'tHooft:1978xw} in our notation and then discuss the
 analytic continuations. The various functions involved and some
 reference formula are summarized below in Appendix C.

 Defining $\rho = \sqrt{-\CK}$, where as before
 \bege
 \CK = \a\b+\b\g+\g\a = -{1\over 4}\big(a^2+b^2+c^2-2(ab+bc+ca)\big),
 \ende
 we have
 \bege\label{JMI3}
 \bar{J_M} = -{1\over 2 \rho}(I_3(a|b,c)+I_3(b|c,a)+I_3(c|a,b)).
 \ende
The results for $I_3(a|b,c)$ can be expressed in terms of the
velocity $v=\sqrt{1-{4(M^2-i\epsilon)\over a}}$ and the variable
$x={\beta/ \rho}$ :
\begin{eqnarray}
 I_3(a|b,c) &=& {\rm Li}_2(z_1) - {\rm Li}_2(\bar{z_1}) + {\rm Li}_2(z_2) - {\rm Li}_2(\bar{z_2})
 -\eta(x-v,x+v)\log{\bar{z_2}\over z_1} \nonumber\\
 &+& \eta(-1-v,{1\over x-v})\log{z_1} - \eta(1-v,{1\over x-v})\log{\bar{z_2}} \nonumber\\
 &+& \eta(-1+v,{1\over x+v})\log{z_2} - \eta(1+v,{1\over x+v})\log{\bar{z_1}}
\end{eqnarray}
where we have defined
 \bege
 z_1 = {x+1\over x-v} \;\;\; z_2 = {x+1\over x+v} \;\;\; \bar{z_1} = {x-1\over x+v} \;\;\;  \bar{z_2} = {x-1\over x-v}
 \ende
 and the function $\eta(x,y)$ compensates for the branch cut in the logarithms:
 \begin{eqnarray}
 \log{xy} &=& \log{x}+\log{y} + \eta(x,y) \\
 \eta(x,y) &=& 2i\pi \Big( \theta(-\Im{x})\theta(-\Im{y})\theta(\Im{x}\Im{y})
 - \theta(\Im{x})\theta(\Im{y})\theta(-\Im{x}\Im{y})\Big)\nonumber.
 \end{eqnarray}

The other two integrals $I_3(b|c,a)$ and $I_3(c|a,b)$ are easily
obtained by permutation of the above results, so that
$x=\gamma/\rho, v=\sqrt{1-{4M^2\over b}}$ and $x=\alpha/\rho,
v=\sqrt{1-{4M^2\over c}}$, respectively. Although these results
entirely characterize the massive triangle function, it is a rather
tedious exercise to analytically continue the results to the six
different physical kinematical regions. To our knowledge, such
complete analytic continuations have not appeared in the literature
thus far.

$\bar{J_M}$ takes different forms for $\CK>0$ and $\CK<0$ since
then the variable $x$ is imaginary and real, respectively. The
case $\CK>0$ can occur only if all momenta are spacelike
$(a,b,c<0)$ or timelike $(a,b,c>0)$. The case $\CK<0$ can occur
for momenta of any signature. Thus, if all momenta are spacelike
or all timelike, the ratios of momenta will determine if $\CK>0$
or $\CK<0$.
 For each of these two cases, we must also distinguish when $a$ is spacelike,
timelike below threshold, and timelike above threshold. For timelike
above threshold and spacelike, the generalized velocity
$v=\sqrt{1-{4M^2\over a}}$ is real, except for the $i\e$ term which
is used in the analytic continuation and hence not included below.
Below threshold $v=i\bar{v}=i\sqrt{{4M^2\over a}-1}$.

 {\center{\Large  $\hspace{2in}  {\rm \bf Case} \;\;\bf{\CK>0}$  }}

 For $\CK>0$ we have
 \begin{eqnarray}
 \rho &=& \sqrt{-\CK} \;=\; i\sqrt{\CK} \;\equiv\; i\bar{\rho} \nonumber\\
 x &=& \b/\rho \;=\; -i\b/\bar{\rho} \;\equiv\; -iw
 \end{eqnarray}
 \begin{itemize}
 \item{$\CK>0 \;\;\;\;{\rm and}\;\;\;\;  v\;{\rm real}  \Longleftrightarrow (a<0 \;{\rm or}\; a>4M^2)$}
 \begin{eqnarray}\label{Kpvre}
  I_3(a|b,c) &=& i \Bigg( 2\Cl2(2\phi_1) - \Cl2(2\phi_1-2\phi_2)-\Cl2(2\phi_1+2\phi_2)\nonumber\\
  &+& 2i\pi(\phi_1-\phi_2)\th(a-4M^2)\Bigg)
  \nonumber\\
  \phi_1 &=& \tan^{-1}(w) \;\;\;\;\;\;\;\;\;\;\; \phi_2 = \tan^{-1}(w/v)
 \end{eqnarray}
 \item{$\CK>0 \;\;\;\;{\rm and}\;\;\;\;   v=i\bar{v} \Longleftrightarrow (0 < a < 4M^2)$}
 \begin{eqnarray}\label{Kpvim}
  I_3(a|b,c) &=& i \Bigg( 2\Cl2(2\bar{\phi}_1) - \Cl2(2\bar{\phi}_1-2\bar{\phi}_2)-\Cl2(2\bar{\phi}_1+2\bar{\phi}_2)
  \nonumber\\
  &+& 2\bar{\phi}_2\log{\left| {w-\bar{v}\over w+\bar{v}} \right|} -2i\pi\bar{\phi}_1\th(|w|-\bar{v}) \Bigg)
 \nonumber\\
  \bar{\phi}_1 &=& \tan^{-1}(1/w) \;\;\;\;\;\;\;\;\;\;\; \bar{\phi}_2 = \tan^{-1}(1/\bar{v})
 \end{eqnarray}
 \end{itemize}
Note that the prefactor of $i$ in the above equations cancels
against the $i$ from $\rho=i\bar{\rho}$ in the prefactor of
Eq.(\ref{JMI3}), so that the terms involving the Clausen function
$\Cl2(x)$ (discussed in Appendix C) contribute to the real part of
$J_M$.

 {\center{\Large  $\hspace{2in} {\rm \bf Case} \;\;\bf{\CK<0}$ }}

Here $x$ is real.
 \begin{itemize}
 \item{$\CK<0 \;\;\;\;{\rm and}\;\;\;\;    v\;{\rm real}  \Longleftrightarrow (a<0\;{\rm or}\; a>4M^2)$}
 \begin{eqnarray}
 I_3(a|b,c) &=& \Re{\Big( \Li2(z_1)-\Li2(\bar{z_1})+\Li2(z_2)-\Li2(\bar{z_2})\Big)}
 \nonumber\\
 &+& 2i\pi\Big( \vp_1 \s(a) \th(|x|-v) + \vp_2\th(a-4M^2) \Big)
 \nonumber\\
 \vp_1 &=& \half \log{\left| {x+1\over x-1} \right|} \;\;\;\;\;\;\;
 \vp_2 = \half \log{\left| {(x+v)(x-1)\over (x-v)(x+1)} \right|}
 \end{eqnarray}
 \item{$\CK<0  \;\;\;\;{\rm and}\;\;\;\;   v=i\bar{v} \Longleftrightarrow (0 < a < 4M^2)$}
 \begin{eqnarray}
 I_3(a|b,c) &=& 2\Re{\Big( \Li2(z_1)-\Li2(\bar{z_1}) \Big)}
 \nonumber\\
 z_1&=& {x+1\over x-i\bar{v}} \;\;\;\;\;\;\;\;\; \bar{z_1} = {x-1\over x+i\bar{v}}
 \end{eqnarray}
 \end{itemize}

Several features of these results deserve comment.

First, in the $\CK>0, v=i\bar{v}$ case, there are anomalous
thresholds which give rise to a nonzero imaginary part and a
diverging real part. As seen in Eq.(\ref{Kpvim}), these anomalous
thresholds occur in $I_3(a|b,c)$ (and similarly for $I_3(b|c,a)$ and
$I_3(c|a,b)$ by permutation) when
 \bege
 |w|=\bar{v} \Longleftrightarrow abc=4M^2\CK.
 \ende
There will be a nonzero imaginary part for $|w|>\bar{v}
\Longleftrightarrow abc>4M^2\CK.$
 Note that since here
$4M^2>a>0$ and $\CK>0$, we must have $b,c>0$. Let us now look at
some special cases:
\begin{itemize}
\item{$\bf{a=b=c}$}\;\;\; Here the condition for an anomalous
threshold reduces to $a>3M^2$, which was found in
\cite{Davydychev:2001uj}.
\item{$\bf{b=c}$}\;\;\; This leads to
$(a/M^2)=(b/M^2)(4-(b/M^2))$, which is possible only if $b<4M^2$.
\end{itemize}
There are also anomalous thresholds for the case of $\CK<0$ and
$a>4M^2$. For example, for the mixed signature symmetric case
$a=b=-c>0$, there is a discontinuity in the real part of
$J_M(a,a,-a)$ and a divergence in the imaginary part at $a=5M^2$, as
seen in Fig.(\ref{fig:9}). Anomalous thresholds were analyzed long
ago \cite{Landau:1959fi}\cite{blanknambu}.

In the case $\CK>0, v\;{\rm real}$, there is an imaginary part above
threshold, $a>4M^2$, which vanishes in the massless limit ${a\over
M^2} \raw \infty$.

In \cite{Davydychev:1997wa}, the authors find an interesting
geometrical interpretation and derivation of the triangle integral
(and higher $n$-point integrals).

In the symmetric limit $a=b=c$, the above results reduce to those
given in Eqs.(55-62) of \cite{Davydychev:2001uj}.

In the massless limit, we obtain
\begin{itemize}
 \item{${\cal K} > 0$}
 \begin{eqnarray}\label{JKp}
 \bar{J}(a,b,c) &=& -{1 \over \rho}\Big( \Cl2(2\phi_\a) + \Cl2(2\phi_\b) + \Cl2(2\phi_\g)\Big)\nonumber\\
 \phi_\a &=& \arctan{\left( \rho\over \a\right)}, \;\; {\rm etc.}
 \end{eqnarray}
 \item{${\cal K} < 0$}
\begin{eqnarray}\label{JKm}
 \bar{J}(a,b,c) &=& -{1 \over \rho}\Bigg( \tilde{\Clh2}(2\phi_\a) + \tilde{\Clh2}(2\phi_\b)
 + \tilde{\Clh2}(2\phi_\g) \nonumber\\
 &+& i\pi\phi_\a \theta(a) + i\pi\phi_\b \theta(b) + i\pi\phi_\g \theta(c)
 \Bigg)\nonumber\\
 \phi_\a &=& \half\log{\left| \a+\rho \over \a-\rho \right|}, \;\; {\rm etc.}
 \end{eqnarray}
where
 \bege
 \tilde{\Clh2}(2\phi_\a) =
 \left( \begin{array}{ccc} \Clh2(2\phi_\a)\;\;\;\;\; {\rm for}\;\;\; ab>0  \\ \AClh2(2\phi_\a)\;\;\;\;\; {\rm for}\;\;\; ab<0
 \end{array}\right),
 \ende
 and similarly for $\tilde{\Clh2}(2\phi_\b)$ when $(bc>0, bc<0)$ and $\tilde{\Clh2}(2\phi_\g)$ when $(ca>0, ca<0)$.
\end{itemize}
The results for the massless case are well known
\cite{'tHooft:1978xw}\cite{Ball}\cite{Davydychev:1996pb}\cite{Lu3gg},
although the notation is non-standard. Here we have adopted the
notation of \cite{Lu3gg} by using the hyperbolic Clausen function
$\Clh2(x)$, and alternating hyperbolic Clausen function
$\AClh2(x)$, which are discussed below.

\vspace{0.75cm}
 {\center{\Large \bf Appendix C: Special Functions}}
\vspace{0.75cm}

 Here we collect some useful results, mainly taken
from \cite{lewin}. The dilogarithm function is defined for complex
$z$ by
 \bege
 {\rm Li}_2(z) = -\int_0^z dx \;{\log{(1-x)\over x}}.
 \ende
 In order to find the real and imaginary parts of this function,
 one should first ensure that the modulus is less than unity by
 judiciously using
 \bege
 \Li2(z) = -\Li2(1/z) -{\pi^2\over 6} -\half \log^2{(-z)}.
 \ende
The notation $\Li2(r,\th)$, with two arguments, is used for the real
part of $\Li2(re^{i\th})$. For modulus less than unity, $r<1$, we
have the integral representation
 \bege
 \Li2(r,\th) = -\half \int_0^r{\log{(1-2x\cos{\th}+x^2)}\over x}
 \ende
 The imaginary part for $r<1$ is
 \begin{eqnarray}
 \Im{(\Li2(r e^{i\theta}))} &=& T\log{r} +\half \left( \Cl2(2\th)+\Cl2(2T)-\Cl2(2\th+2T)
 \right)\nonumber\\
 T&=&\tan^{-1}{\left( {r\sin{\th}\over 1-r\cos{\th}} \right)}.
 \end{eqnarray}
In particular,
 \bege\label{Cl2def}
 \Im(\Li2(e^{i\th})) = \Cl2(\th)\;\;\;\;\;{\rm and}\;\;\;\;\;
 \Cl2(\th) = {1\over 2i}\left( \Li2(e^{i\th})-\Li2(e^{-i\th})\right).
 \ende
The Clausen function frequently appears in the triangle integral
and has the following representations :
 \bege
 \Cl2(x) = -\int_0^x dy\log{|2\sin{y\over 2}|} =
 \sum_{n=1}^\infty{\sin{nx}\over n^2} \nonumber\\
 \ende
 Furthermore, $\Cl2(x)$ is odd, $\Cl2(-x)=-\Cl2(x)$, satisfies periodicity,
 $\Cl2(x+2n\pi)=\Cl2(x)$, and a duplication formula
 $\Cl2(2x) = 2\Cl2(x) + 2\Cl2(x-\pi)$. Many other properties can be
 found in \cite{lewin} and the some are conveniently summarized in
 the appendix of \cite{Lu3gg}.

We have used the notation of Lu\cite{Lu3gg}, who used the
hyperbolic Clausen function, $\Clh2(x)$, and alternating
hyperbolic Clausen function, $\AClh2(a)$, defined by the integral
representations
 \begin{eqnarray}
 \Clh2(x) &=& -\int_0^x dy\log{|2\sinh{y\over 2}|}
 \nonumber\\
 \AClh2(x) &=& -\int_0^x dy\log{|2\cosh{y\over 2}|}.
 \end{eqnarray}
 These can also be written as
\begin{eqnarray}
 \Clh2(x) &=& \half\Re{\!\left( \Li2(e^x)-\Li2(e^{-x}) \right)}  \nonumber\\
  \AClh2(x) &=& \half\Re{\!\left( \Li2(-e^x)-\Li2(-e^{-x}) \right)}
\end{eqnarray}
in analogy with Eq.(\ref{Cl2def}).

 Finally, some elementary relations which are used often include
 (for $x,y$ real)
 \begin{eqnarray}
 \arg{(x+iy)} = \tan^{-1}\left({y\over x}\right) +\pi \th(-x)\s(y) \nonumber\\
 \tan^{-1}(x)+\tan^{-1}(y) = \tan^{-1}\left({x+y\over 1-xy}\right) + \pi
 \s(x)\th(xy-1)\nonumber\\
\tan^{-1}(x)+\tan^{-1}\left(1/x\right) = \s(x){\pi\over 2}.
 \end{eqnarray}
where $\s(x)= x/|x|$ is the sign function and the step function
$\th(x)=(\s(x)+1)/2$ should not be confused with the angle $\th$.

\vspace{0.75cm}
 {\center{\Large \bf Appendix D: Form Factors with Dimensional Reduction (DRED) Regularization}}
\vspace{0.75cm}

Here we discuss the form factors regularized using dimensional
reduction (DRED) in integer number of dimensions $d_R$, defined
analogously to the usual $d_R=4$ DRED scheme.  This could be used
for $d_R=6$ or $d_R=10$ theories, but of course we mainly have in
mind the four-dimensional case.

It is easy to see that the quark and scalar contributions are
unchanged from DREG, and only the gluon contribution is different.
This is most easily expressed in terms of the modified sum rule
 \bege
 F_G(DRED)+4F_Q+(10-d_R)F_S=0,
 \ende
which implies $F_G(DRED)=F_G(DREG) + (d_R-d)F_S$. Expanding
$d=d_R-2\e$ around the real number of dimensions $d_R$ leads to
$F_G(DRED)=F_G(DREG) + 2\e F_S$, which makes manifest the role of
the $2\e$ adjoint DRED ``ghosts" which preserve supersymmetry.

 In four dimensions we have
 \bege
 F_G(DRED)+4F_Q+6 F_S=0.
 \ende

Since only the $A$ form factors have a UV divergence in four
dimensions, only these form factors will be changed when using DRED:
 \bege
 \d_{DRED}(A_{12}(G)) = \d_{DRED}(\bar{A}_{12}(G)) = -{1\over 3} {i\over
 16\pi^2}.
 \ende
 In the symmetrized physical $\pm$ basis we have
 $\d_{DRED}(A_0(G)) = -{1\over 3} {i\over 16\pi^2} $, and all other
 form factors are unchanged.

 \vspace{0.75cm}
 {\center{\Large \bf Appendix E: Quark and Squark Mass Corrections }}
\vspace{0.75cm}

Here the corrections to the form factors due to fermion and scalar
masses will be given. The massive quark (MQ) contributions were
first obtained in \cite{Davydychev:2001uj}, and we obtain exactly
the same results. To our knowledge, the squark contributions,
either massless or massive (MS), have not yet appeared in the
literature.

First, the well known formulas for the scalar and fermion
self-energies are reproduced in our notation :
 \begin{eqnarray}\label{PiM}
 \Pi_1(MQ)&=& {d-2\over 1-d}  J_{1M} +2M^2\Bigg( {2J_{1M}-(d-2)T_M\over a(1-d)} \Bigg)
 \nonumber\\
 \Pi_1(MS)&=& {1\over 2(1-d)} J_{1M}- M^2\Bigg( {2J_{1M}-(d-2)T_M\over a(1-d)} \Bigg)
 \end{eqnarray}
 with the integrals $J_{1M},T_M$ given in Appendix A.
These yield the massive fermion and scalar contributions to the
longitudinal form factors through Eq.(\ref{LFFs}).

Notice that the terms proportional to $M^2$ in the above two
equations are equal up to a factor of $-2$. After explicit
calculation, it was discovered that the scalar mass correction terms
($\d_{MS}$) are just minus one-half of the quark mass correction
terms ($\d_{MQ}$) for all form factors. Thus, for generic form
factor $F$
 \begin{eqnarray}\label{dMQS}
 F(MS) &=& F(S)\Big|_M + \d_{MS}(F)\nonumber\\
 F(MQ) &=& F(Q)\Big|_M + \d_{MQ}(F)\\
 \d_{MQ} &=& -2\d_{MS} \nonumber
 \end{eqnarray}
The notation $F(S)\Big|_M$ simply means to take the appropriate
massless result for the form factor $F$, as given in section 3, and
replace $J\raw J_M$, $J_1\raw J_{1M}$, $J_2\raw J_{2M}$, $J_3\raw
J_{3M}$ everywhere.

Because of the relation $\d_{MQ} = -2\d_{MS}$, we need only write
either the fermion or scalar mass correction terms explicitly. Here
we choose the scalar contributions, which for the transverse form
factors are
 \begin{eqnarray}
 \delta_{MS}(\bar{F}_{12}) &=& -{2M^2\over \CK^2} \Bigg[ \left(
 {\CK-3\b\g\over d-2}\right) J_M +{\CP-2\a\CK-\g^2(3\a+2\g-\b)\over
 a(\b-\g)(d-1)}J_{1M}\\
 &-&{\CP-2\a\CK-\b^2(3\a+2\b-\g)\over
 b(\b-\g)(d-1)}J_{2M} -{2c\over d-1}J_{3M}
 + {(2-d)\CK\a\over 2(d-1)ab}T_M \Bigg] \nonumber
 \end{eqnarray}
and
\begin{eqnarray}
 \delta_{MS}(\bar{H}) = {2M^2\over \CK^2} \Bigg[ {3\CP\over d-2} J_M -{\CK-2\a\g\over d-1}J_{1M}
 -{\CK-2\a\b\over d-1}J_{2M}-{\CK-2\b\g\over d-1}J_{3M}
 +{\CK(d-2)\over 2(d-1)}T_M \Bigg].
 \end{eqnarray}

The results in the physical $\pm$ basis can be obtained from those
of the $LT$ basis through the use of Eq.(\ref{pmLTrel}), and are
included here for completeness.

 \begin{eqnarray}\label{A12M}
  \d_{MS}(A_{12}) = {M^2\over \CK} \Bigg[ {c\a\over d-2}J_M
 + {\g J_{1M} +\b J_{2M} +c J_{3M} \over d-1} \Bigg]
 \end{eqnarray}
\begin{eqnarray}
 \d_{MS}(B_{12}) =-{M^2\over \CK} \Bigg[ {(a-b)\a\over d-2}J_M
 + {(2\a+\g)J_{1M} - (2\a+\b)J_{2M} +(\b-\g)J_{3M} \over d-1}
 \Bigg]
 \end{eqnarray}
 \begin{eqnarray}
 \d_{MS}(C_{12}) &=& {M^2\over 4\CK^2}
 \Bigg[ {8\CP +\a^2c+3\b^2(\a-\g)+3\g^2(\a-\b)\over d-2}J_M \nonumber\\
 &+& {8\CP +\a^2(\b-\g)+\g^2(5\a+7\b+4\g)\over a(d-1)}J_{1M}\\
 &+& {8\CP +\a^2(\g-\b)+\b^2(5\a+7\g+4\b)\over b(d-1)}J_{2M} \nonumber\\
 &+& {4c^2+2\b\g-\CK\over d-1}J_{3M} + {(d-2)\CK(\a^2+2\b\g+3\CK)\over (d-1)2ab}T_M \Bigg]
 \nonumber
 \end{eqnarray}
\begin{eqnarray}
\d_{MS}(D_{12}) &=& {M^2\over 4\CK^2}
 \Bigg[ {(\CK+3\a^2)(a-b)\over d-2}J_M + {\CK+4\a^2+2\a\g\over d-1}J_{1M} \\
 &-&  {\CK+4\a^2+2\a\b\over d-1}J_{2M} +{(2\b\g-3\CK)(a-b)\over c(d-1)}J_{3M}
 + {(d-2)\CK(a-b)\over 2(d-1)c}T_M \Bigg]
\nonumber
\end{eqnarray}
\begin{eqnarray}
 \d_{MS}(H) &=& {M^2\over 4\CK^2}
 \Bigg[ {2\CP+abc\over 2-d}J_M +{\CK-2\a\g\over d-1}J_{1M}
 + {\CK-2\a\b\over d-1}J_{2M} +{\CK-2\b\g\over d-1}J_{3M} \nonumber\\
 &-& {(d-2)\CK\over 2(d-1)}T_M \Bigg)\Bigg]
 \end{eqnarray}
\begin{eqnarray}
\d_{MS}(S) &=& -{M^2\over 4\CK^2}
 \Bigg[ {3(a-b)(b-c)(c-a)\over 2-d}J_M + {(4a^2+2\a\g-3\CK)(b-c)\over a(d-1)}J_{1M}
  \nonumber\\ &+&  {(4b^2+2\a\b-3\CK)(c-a)\over b(d-1)}J_{2M}
  + {(4c^2+2\b\g-3\CK)(a-b)\over c(d-1)}J_{3M} \nonumber\\
  &-& {(d-2)(a-b)(b-c)(c-a)\CK\over 2(d-1)abc}T_M \Bigg)\Bigg].
\end{eqnarray}

 It is straightforward to see that all of the correction
terms are ultraviolet finite.

The relation $-2\d_{MS}=\d_{MQ}$ is necessary for the preservation
of the form of a quark/scalar sum $\Sigma_{SQ} = 2F_S+F_Q$, which is
equal to ${2\over d-10}\Sigma_{QG}$ using the results of section 3.
In other words $2F_{MS}+F_{MQ} = 2F_S\big|_M+F_Q\big|_M$, so that
this quantity has no correction terms proportional to $M^2$.

However, the relations between massive gauge bosons and massive
fermions and/or scalars will be different, since the gauge bosons
eat a degree of freedom to acquire mass. Consider the contribution
of a massive gauge boson to the gauge-invariant gluon self-energy
\footnote{This is also the contribution of $W^{\pm}$ to the photon
self-energy.} :
 \begin{eqnarray}
 \Pi_1(MG) &=& J_{1M}\left[ {8d-d_R-7\over 2(d-1)} \right]
 + {2M^2\over a}{(d_R-1)\over (d-1)}\left[ J_{1M}+ \half (2-d)T_M
 \right],
 \end{eqnarray}
 where as before $d_R=d$ in dimensional regularization (DREG) and
 $d_R=4$ (or the real integer number of dimensions) in dimensional
 reduction (DRED).
 From this and Eq.(\ref{PiM}) we deduce
 \bege
 \Pi_1(MG) + 4\Pi_1(MQ) +(9-d_R)\Pi_1(MS) =0
 \ende
and thus the massive $N=4$ sum rule becomes
 \bege
 F_{MG} + 4F_{MQ} +(9-d_R)F_{MS} =0
 \ende
for the longitudinal form factors. It can also be shown that this
holds for the transverse form factors and so the results of this
paper also give the contributions of massive internal gauge bosons.
Proving this involves detailed analysis of the vertices and diagrams
that contribute to the triple gluon vertex when the PT/BFMFG is
applied to a spontaneously broken gauge theory that leaves a
non-abelian subgroup (of gluons) intact. This can be done following
a pinch-technique route similar to
\cite{Papavassiliou:1989zd}\cite{Papavassiliou:1993ex}. Due to the
equivalence of the PT and BFMFG, it is more convenient to follow the
BFMFG route similar to \cite{Denner:1994xt}.
 For example, in $SU(5)$
GUTs the colored superheavy $X$ and $Y$ gauge bosons give a
contribution which satisfies the above massive sum rule. We should
emphasize that these sum rules are simply a convenient way of
relating the contributions of various spin particles, all with
hypothetical mass $M$, but no assumption is made about the actual
masses for a given theory under consideration; the sum rules are
entirely stripped of color factors.






\newpage

\end{document}